\documentclass{article}
\usepackage{graphicx} 
\usepackage{subcaption}
\usepackage{multirow}
\usepackage{rotating}
\usepackage{bm}
\usepackage{amsmath}
\usepackage{makecell}
\usepackage{booktabs}
\usepackage{hyperref}
\usepackage[margin=2cm]{geometry}

\title{GLOW-FDG: Generalized cancer LesiOn Whole-body segmentation model for $^{18}$F-FDG-PET/CT}

\author{
Maksym Fritsak$^{1,2,\dagger,*}$ \and
Maximilian Rokuss$^{3,\dagger,*}$ \and
Hubert S. Gabryś$^{1}$ \and
Yannick Kirchhoff$^{3}$ \and
Benjamin Hamm$^{3}$ \and
Sebastian M. Christ$^{1}$ \and
Nicolas Martz$^{1,4}$ \and
Isabelle Opitz$^{5}$ \and
Rolf Stahel$^{6}$ \and
Martin Hüllner$^{7}$ \and
Matthias Guckenberger$^{1}$ \and
Klaus Maier-Hein$^{3,\ddagger}$ \and
Stephanie Tanadini-Lang$^{1,\ddagger}$
}

\date{}

\begin{document}
\maketitle

\begin{center}
\small

$^{1}$ Department of Radiation Oncology, University Hospital Zurich, University of Zurich, Zurich, Switzerland\\
$^{2}$ Faculty of Medicine, University of Zurich, Zurich, Switzerland\\
$^{3}$ Division of Medical Image Computing, German Cancer Research Center (DKFZ), Heidelberg, Germany\\
$^{4}$ Department of Radiation Oncology, Institut de Cancérologie de Lorraine, Vandœuvre-lès-Nancy, France\\
$^{5}$ Department of Thoracic Surgery, University Hospital Zurich, University of Zurich, Zurich, Switzerland\\
$^{6}$ ETOP IBCSG Partners Foundation, Bern, Switzerland\\
$^{7}$ Department of Nuclear Medicine, University Hospital Zurich, University of Zurich, Zurich, Switzerland\\[1em]

$^{\dagger}$ These authors contributed equally to this work.\\
$^{\ddagger}$ These authors jointly supervised this work.\\[0.5em]

$^{*}$ Corresponding authors:
\texttt{maksym.fritsak@usz.ch},
\texttt{maximilian.rokuss@dkfz-heidelberg.de}

\end{center}

\section*{Abstract}
Whole-body fluorodeoxyglucose positron emission tomography combined with computed tomography is widely used in cancer care, but manual lesion delineation is slow, subjective, and difficult to scale. We present GLOW-FDG, an open-source artificial intelligence model for whole-body cancer lesion segmentation in fluorodeoxyglucose positron emission tomography and computed tomography. The model was trained on 1,563 scans spanning multiple cancer types and evaluated on 185 external scans from independent institutions. Across breast cancer, nonmetastatic and oligometastatic lung cancer, head and neck cancer, and metastatic melanoma, GLOW-FDG consistently outperformed publicly available benchmark models in lesion detection, while reducing false positives and maintaining strong segmentation accuracy. Quantification of total tumor burden and total lesion glycolysis was robust across cohorts, and performance approached the variability observed between expert radiation oncologists. These results support GLOW-FDG as a generalizable tool for automated cancer segmentation and quantitative imaging biomarker extraction in whole-body imaging.

\section{Introduction}

Whole-body $^{18}$F-fluorodeoxyglucose positron emission tomography combined with computed tomography (FDG-PET/CT) is widely adopted in oncologic staging \cite{Cuaron2012FDGPETNSCLC, Fletcher2008FDGPETOncology}, response assessment \cite{Cuaron2012FDGPETNSCLC, Wahl2009PERCIST}, and prognostication \cite{Cuaron2012FDGPETNSCLC, Hildebrandt2022FDGPETCTBreast} due to its visualization of tumor metabolism against anatomic context. Quantitative, volumetric biomarkers derived from FDG-PET, such as metabolic tumor volume (MTV), disease dissemination indices, and total lesion glycolysis (TLG), carry independent prognostic value \cite{Pak2014MTVTLGHNC, Cottereau2020DLBCLDissemination} and are increasingly considered for risk stratification and therapy guidance \cite{Hirata2021QuantitativeFDGPET, Marcus2025FDGPETHNSCC}. However, manual cancer lesion delineation is labor-intensive and subject to inter-observer variability, limiting routine clinical deployment and large-scale studies \cite{HATT2018177}, thereby motivating reliable automation. 

Artificial intelligence (AI) has rapidly advanced medical image segmentation, particularly for anatomically consistent targets like organs-at-risk \cite{wasserthal2023totalsegmentator, xu2025cadscomprehensiveanatomicaldataset}, enabling streamlined pipelines, reduced manual effort, and improved consistency \cite{Lucido2023DLAutoSegOAR}. Yet cancer lesion segmentation is fundamentally harder: cancer lesions vary widely in shape, size, and do not obey the anatomic constancy of normal organs. Tumor FDG uptake can also overlap with numerous physiologic or inflammatory processes; as a result, models trained on narrow cohorts of data often generalize poorly when presented with whole-body images or previously unseen cancers \cite{Faro2026DLTumorBurdenMelanoma}. These limitations hinder clinical utility and cross-study comparability.
 
Another source of error is intra-observer variability in reference annotations \cite{Yang2020ContouringRadiomics}, which increases label noise and complicates both model training and validation. Large public datasets have helped to mitigate this issue, but they typically emphasize only a subset of cancers or body regions \cite{Gatidis2022FDGPETCT, Oreiller2022HECKTORChallenge}. Consequently, the need for a comprehensive dataset that covers multiple cancer types in a whole-body imaging setting remains unmet. Both academic and industry perspectives suggest that automated FDG-PET/CT whole-body tumor segmentation is moving from research into clinical use. Commercial tools often combine thresholding with AI models to distinguish pathology from normal physiology \cite{Faro2026DLTumorBurdenMelanoma, ShiyamSundar2024}, while academic research is converging on deep learning methods, particularly U-Net-based architectures \cite{rokuss2024fdgpsmahitchhikersguide, kalisch2024autopetiiichallengeincorporating}. 
Although existing lesion segmentation models, including those approaching interrater performance, predominantly rely on prompts like points, boxes or text~\cite{Rokuss_2025_CVPR,isensee2025nninteractiveredefining3dpromptable,r2025voxtell} or are restricted to CT~\cite{GF-Screen,lei2026syntheticdatadrivenradiologyfoundation}, existing models have shown limited performance for FDG-PET/CT tumor segmentation \cite{fritsak2026generalizing}, where lesion size, physiologic uptake, PET–CT misalignment, and heterogeneous lesion morphology introduce additional challenges.

A critical gap that remains is the absence of a publicly available FDG-PET/CT cancer lesion segmentation model trained at large scale on a highly diverse, multi-cancer curated dataset and validated on separate, independent external cohorts. Without both diversity in training and independence in validation, reported gains are vulnerable to site- or cohort-specific effects rather than genuine robustness, limiting clinical credibility and regulatory readiness.

In this study, we present an open-source whole-body FDG-PET/CT cancer lesion segmentation model, GLOW-FDG, trained on a diverse curated corpus spanning lung cancer, head and neck cancer, prostate cancer, lymphoma, melanoma, and soft tissue sarcoma, including more than $1500$ FDG-PET/CT images and evaluated on $185$ held-out scans from other institutions. Building on the evidence above, our design emphasizes (i) data augmentation to account for misalignment between PET and CT scans, (ii) organ supervision as multitask learning to reduce false positives and improve generalization, and (iii) loss function adjustments that stabilize training and balance lesion versus organ segmentation objectives. 

Beyond introducing and evaluating GLOW-FDG, this study systematically compares its performance with several publicly available FDG-PET/CT segmentation models. We benchmark lesion segmentation and lesion detection performance across diverse, independent external validation cohorts to assess robustness and generalization. Additionally, to contextualize model performance relative to human experts, we evaluated the segmentation and detection performance of two radiation oncologists on a small subset of the metastatic melanoma test data to provide insight into the quality of model predictions in relation to expert-level performance.

\section{Results}

Validation of GLOW-FDG was performed on five independent FDG-PET/CT cohorts covering breast cancer, nonmetastatic and oligometastatic lung cancer, head and neck cancer, and metastatic melanoma. Model performance was evaluated for lesion detection and segmentation and compared against three publicly available benchmark methods: the AutoPET III winning model from DKFZ (AutoPET DKFZ) \cite{rokuss2024fdgpsmahitchhikersguide}, the FDG-PET/CT model from IKIM (AutoPET IKIM) \cite{kalisch2024autopetiiichallengeincorporating}, and the PET-only model onlyPET \cite{salimi_lesion_segmentation}. The results are presented first for patient- and lesion-level detection across all validation cohorts, followed by group-wise error analysis for GLOW-FDG, patient- and lesion-wise segmentation analyses, and comparison of inter-observer variability between two radiation oncologists.

\subsection{Patient- and Lesion-wise Detection: Model Comparison}
Across all covered cancer types, the proposed GLOW-FDG model consistently achieved the highest patient- and lesion-wise detection performance (Fig. \ref{fig:pat_f1}, Tables \ref{tab:lesion_detection}, \ref{tab:pat_prec}-\ref{tab:pat_f1}).

In breast cancer, GLOW-FDG achieved the highest patient-wise mean F1 score (0.96), outperforming existing models AutoPET DKFZ (0.86), AutoPET IKIM (0.70), and onlyPET (0.93). Recall was uniformly high across all models (0.96–1.00); however, lower F1 values for the benchmark models resulted from reduced precision despite perfect sensitivity. Furthermore, GLOW-FDG demonstrated the narrowest 95\% CI for F1 compared with the broader and lower 95\% CIs of the other models (Table \ref{tab:pat_f1}).

In nonmetastatic lung cancer, GLOW-FDG again demonstrated the strongest performance, with a mean F1 score of 0.98 compared to 0.88 for both AutoPET DKFZ and AutoPET IKIM, and 0.87 for onlyPET. While recall was high across all models (0.95–1.00), reduced precision (0.82–0.85) in the benchmark models lowered their mean F1 scores. The 95\% CI for GLOW-FDG’s F1 score was again narrower and higher than in benchmark models (Table \ref{tab:pat_f1}).

In oligometastatic lung cancer, GLOW-FDG achieved the highest mean F1 score of 0.86, outperforming AutoPET DKFZ with 0.79, as well as AutoPET IKIM and onlyPET both with 0.74. Although recall values were comparable across methods (0.88–0.91), the superior F1 score of GLOW-FDG reflects a more favorable precision–recall trade-off, with a precision of 0.88 compared to 0.67–0.75 for the other models (Table \ref{tab:pat_prec}).

In head and neck cancer, GLOW-FDG achieved a mean F1 score of 0.89, outperforming AutoPET DKFZ with 0.42 and AutoPET IKIM with 0.59, and slightly exceeding onlyPET with 0.86. Recall was identical for GLOW-FDG and onlyPET at 0.90 (Table \ref{tab:pat_recall}).

In metastatic melanoma, GLOW-FDG provided the strongest detection performance with a mean F1 score of 0.83, compared with 0.72 for AutoPET DKFZ, 0.63 for AutoPET IKIM, and 0.56 for onlyPET. Recall was also highest for GLOW-FDG at 0.85, whereas onlyPET underperformed substantially at 0.61. The confidence interval for GLOW-FDG (0.71–0.91) indicated stable detection capacity, while the wider and lower intervals for AutoPET DKFZ (0.58–0.83), AutoPET IKIM (0.48–0.78), and onlyPET (0.41–0.70) emphasized their reduced reliability (Table \ref{tab:pat_f1}).

At the lesion level (Table \ref{tab:lesion_detection}), the GLOW-FDG model reached the highest detection precision in every cohort, with values between 0.83 and 0.98, while maintaining strong recall (0.80–0.98), resulting in the best overall F1 scores (0.82–0.97). In contrast, other models showed lower and more variable performance, with reduced precision leading to notably lower F1 scores (0.48–0.84).

\subsection{Group-wise Analysis: GLOW-FDG}  
Compared with other models (Tables \ref{tab:lesion_detection}, \ref{tab:pat_prec}), GLOW-FDG demonstrated a remarkably low rate of FP lesions (Table \ref{tab:conf}). For oligometastatic lung cancer, our model correctly identified 95 TP lesions, while missing 15 FN lesions, and erroneously identifying 19 FP lesions. In head and neck cancer, 117 TP lesions were detected, with 17 FN and 13 FP lesions. For metastatic melanoma, 70 TP lesions were detected, with 18 FN and 13 FP lesions. Quantitative characterization of TP, FN, and FP groups (\ref{tab:p_lung}–\ref{tab:p_melanoma}) showed that FP lesions generally exhibited smaller volumes and lower glycolysis compared with FN lesions. Statistically significant differences between the respective distributions were in oligometastatic lung cancer for both lesion volume (p $<$ 0.001) and glycolysis (p = 0.004), as well as in metastatic melanoma for lesion volume (p = 0.002) and glycolysis (p = 0.014). Differences in $SUV_{BW}$ between FP and FN lesions were significant for maximum and mean $SUV_{BW}$ in head and neck cancer (p = 0.041 and p = 0.002, respectively) and in lung cancer (p = 0.046 and p = 0.003, respectively).

When comparing TP delineated by AI versus GT (Tables \ref{tab:p_lung}-\ref{tab:p_melanoma}), most metrics showed no significant differences, indicating a close match between AI and manual segmentation. However, exceptions were observed: in metastatic melanoma, both lesion volume (p = 0.045) and mean $SUV_{BW}$ (p = 0.002) differed between TP(GT) and TP(AI), while in head and neck cancer, a significant difference was found for mean $SUV_{BW}$ (p $<$ 0.001). In contrast, lesion glycolysis consistently showed no significant difference between TP(GT) and TP(AI) across all cohorts (p $>$ 0.2).

\subsection{Patient-wise Segmentation: Model Comparison}

\begin{figure}[htbp]
    \centering
    \begin{subfigure}[b]{0.45\textwidth}
        \centering
        \includegraphics[width=\textwidth]{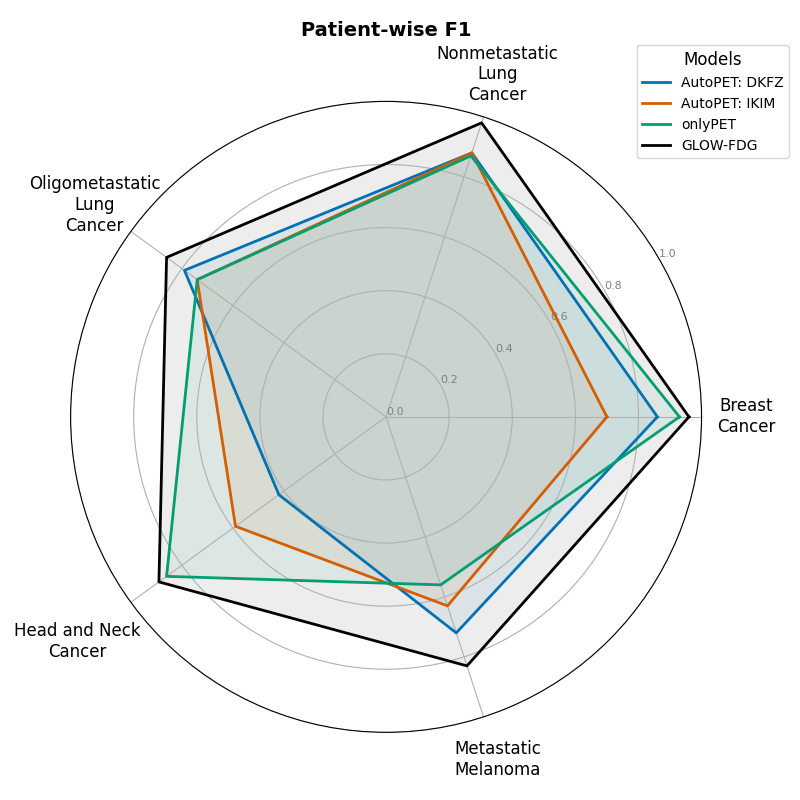}
        \caption{Patient-wise F1}
        \label{fig:pat_f1}
    \end{subfigure}
    \hfill
    \begin{subfigure}[b]{0.45\textwidth}
        \centering
        \includegraphics[width=\textwidth]{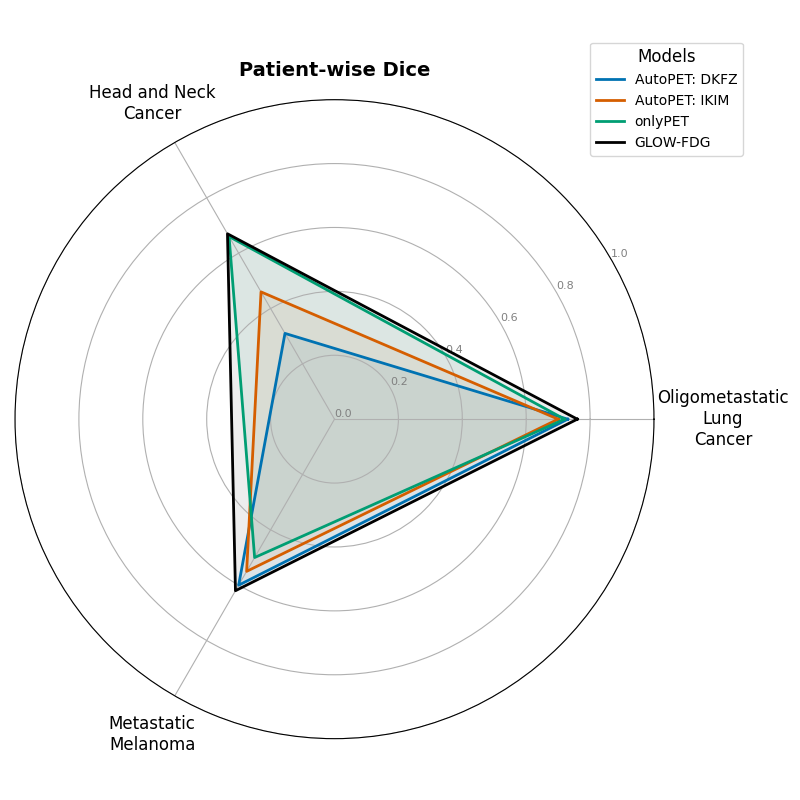}
        \caption{Patient-wise Dice}
        \label{fig:pat_dice}
    \end{subfigure}
    
    \caption{A) Patient-wise F1 score and B) Dice similarity score. Nonmetastatic lung cancer and breast cancer datasets were initially segmented using a pretrained model for 3D semantic segmentation of FDG-avid lesions from PET/CT scans, as part of the segmentations produced by BAMF under AIMI \cite{van_oss_2024_bamf}, and subsequently validated by radiation oncologists from our team to ensure that the cancer lesions were contained within the segmented regions. Only human-segmented datasets were used for Dice score calculation.}
\end{figure}

For oligometastatic lung cancer on the patient level, GLOW-FDG provided nearly unbiased volumetric quantification (Table \ref{tab:pat_wise_GLOW}). Median RPD was –0.5\% for TTB and 4.5\% for TLG. Median ARPD values were 20.8\% for TTB and 10.9\% for TLG, with excellent reproducibility (ICC 0.96 for TTB and 0.99 for TLG). In contrast, AutoPET DKFZ and AutoPET IKIM systematically overestimated tumor burden (Tables \ref{tab:pat_wise_AutoPET_DKFZ}, \ref{tab:pat_wise_AutoPET_IKIM}). The onlyPET model (Table \ref{tab:pat_wise_onlyPET}) showed slight underestimation with median RPD values of –9.7\% for TTB and –0.3\% for TLG although reproducibility remained lower than for GLOW-FDG (ICC 0.92 for TTB and 0.97 for TLG). Dice analysis (Fig. \ref{fig:pat_dice}, Table \ref{tab:pat_Dice}) confirmed these trends: GLOW-FDG achieved the highest overlap (mean Dice 0.78), followed by onlyPET (0.76), AutoPET DKFZ (0.72), and AutoPET IKIM (0.71).

In head and neck cancer (Tables \ref{tab:pat_wise_GLOW}–\ref{tab:pat_wise_onlyPET}), patient-wise analysis for GLOW-FDG showed a median RPD of –27.4\% for TTB and –8.8\% for TLG. Median ARPD values were 29.5\% for TTB and 16.8\% for TLG, with strong reproducibility (ICC 0.79 for TTB and 0.96 for TLG). AutoPET DKFZ substantially underestimated TTB and TLG, while reproducibility was limited (ICC 0.61 for TTB and 0.70 for TLG). AutoPET IKIM showed smaller underestimation, and moderate reproducibility (ICC 0.71 and 0.88). The onlyPET model underestimated less strongly, although reproducibility remained lower than for GLOW-FDG (ICC 0.63 for TTB and 0.89 for TLG). Dice analysis (Fig. \ref{fig:pat_dice}, Table \ref{tab:pat_Dice}) confirmed these findings, with GLOW-FDG achieving the highest mean Dice (0.71), followed by onlyPET (0.68), and AutoPET IKIM (0.52), while AutoPET DKFZ (0.29) performed substantially worse.

In metastatic melanoma (Tables \ref{tab:pat_wise_GLOW}–\ref{tab:pat_wise_onlyPET}), patient-level analysis showed that GLOW-FDG achieved median RPD values of –22.3\% for TTB and –9.9\% for TLG. Corresponding median ARPD values were 27.6\% and 14.2\%, with excellent reproducibility (ICC 0.98 for TTB and 1.00 for TLG). AutoPET DKFZ showed smaller median bias but broader variability, while ICCs remained high (0.92 and 0.99). AutoPET IKIM yielded more variable results with good ICCs of 0.95 and 0.99. The onlyPET model underestimated more strongly with lower reproducibility (ICC 0.90 for TTB and 0.97 for TLG). Dice analysis (Fig. \ref{fig:pat_dice}, Table \ref{tab:pat_Dice}) showed comparable overlap for GLOW-FDG and AutoPET DKFZ (mean Dice 0.70), although GLOW-FDG demonstrated slightly lower variability and a narrower 95\% CI. AutoPET IKIM performed slightly worse (0.67), while onlyPET reached the lowest Dice (0.62).

\subsection{Lesion-wise Segmentation: GLOW-FDG}
\begin{figure}[ht]
\centering
\includegraphics[width=0.6\textwidth]{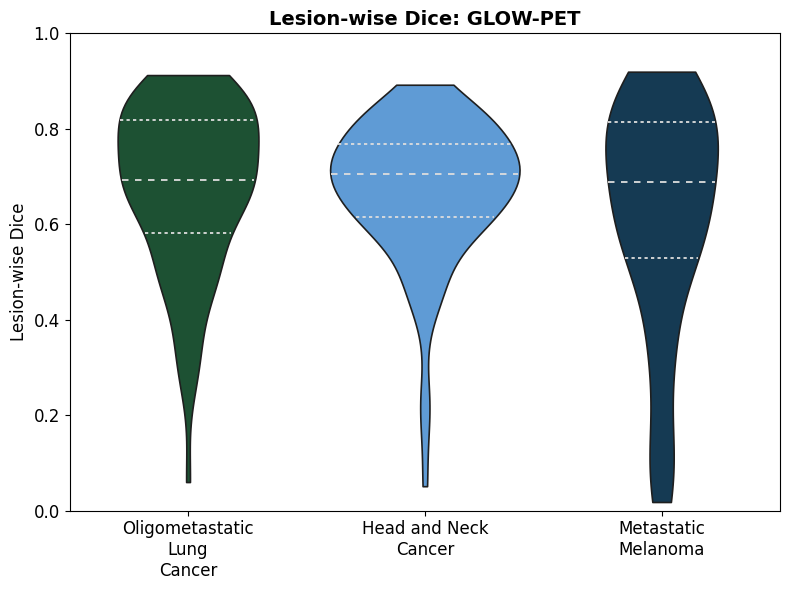}
\caption{GLOW-FDG: lesion-wise Dice. Only human-segmented datasets were used for Dice score calculation.}
\label{fig:glow_fdg_dice}
\end{figure}

\begin{figure}[ht]
\centering
\includegraphics[width=\textwidth]{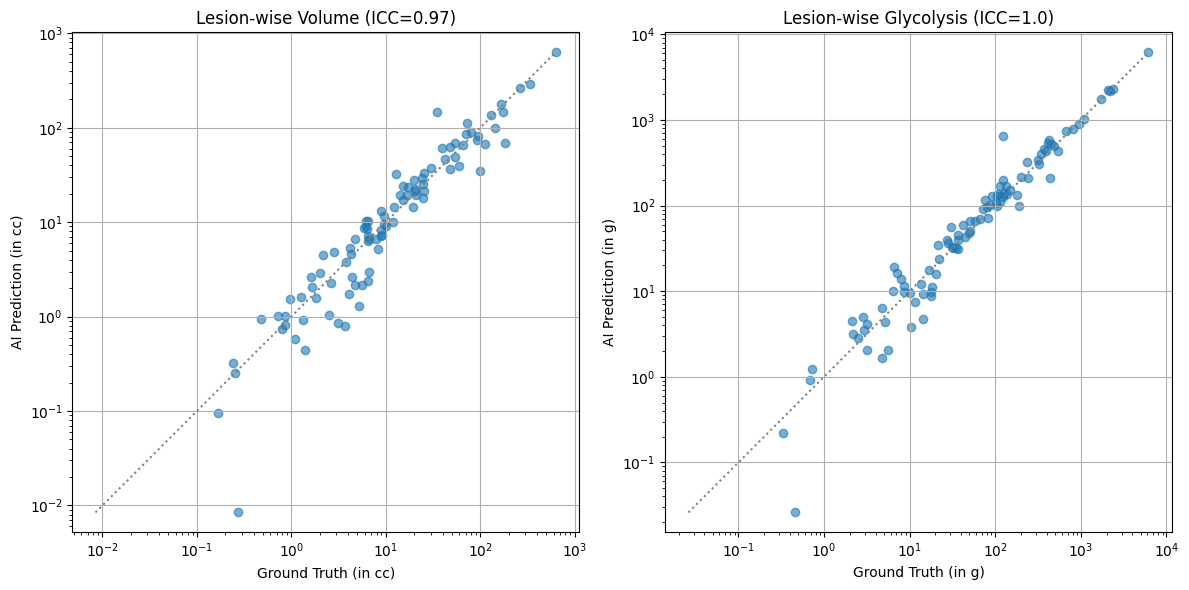}
\caption{Whole-body oligometastatic lung cancer: AI–GT comparison of lesion-wise volume and glycolysis.}
\label{fig:lung}
\end{figure}

\begin{figure}[ht]
\centering
\includegraphics[width=\textwidth]{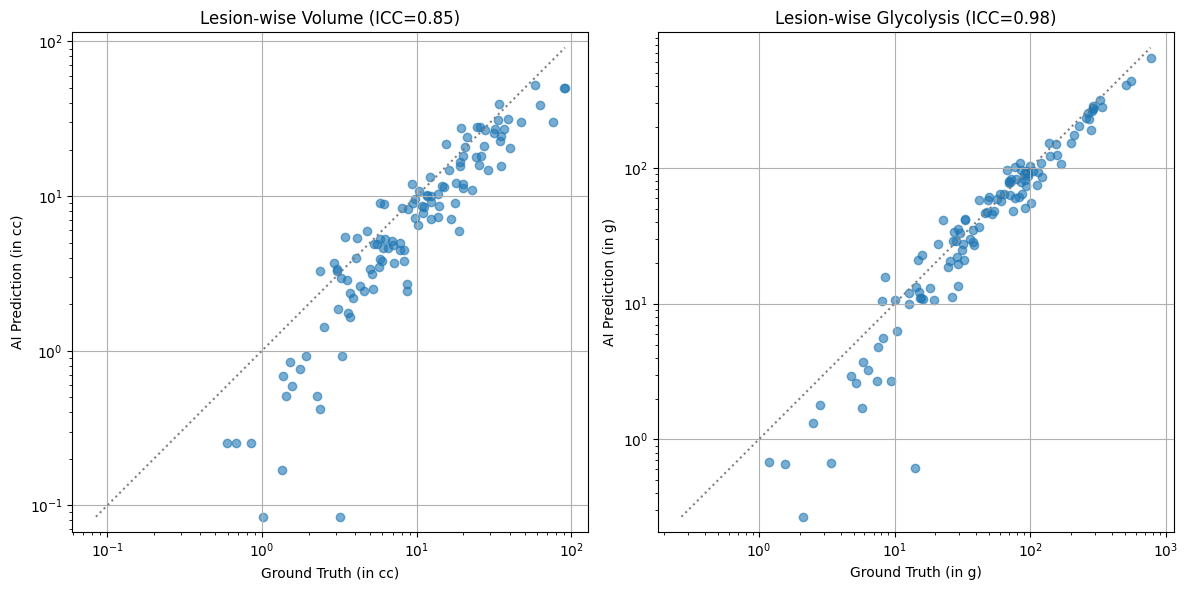}
\caption{Head and neck cancer: AI–GT comparison of lesion-wise volume and glycolysis.}
\label{fig:hn}
\end{figure}

\begin{figure}[ht]
\centering
\includegraphics[width=\textwidth]{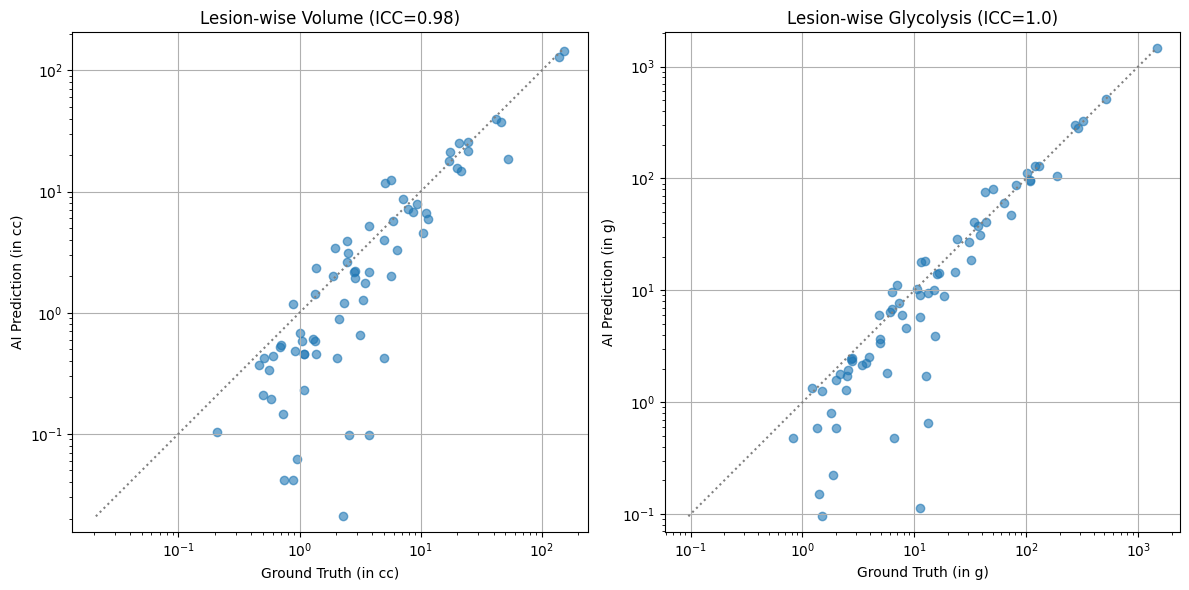}
\caption{Whole-body metastatic melanoma: AI–GT comparison of lesion-wise volume and glycolysis.}
\label{fig:melanoma}
\end{figure}

Overall, the GLOW-FDG model achieved consistent lesion-wise segmentation performance across all three cancer cohorts (Table \ref{tab:glow_fdg_vol_lg}, Fig. \ref{fig:glow_fdg_dice}, Fig. \ref{fig:lung}-\ref{fig:melanoma}). 

In oligometastatic lung cancer, the model achieved a median Dice score of 0.69, lesion volumes were accurately reproduced, with a near-zero median RPD of 0.4\% and a median ARPD of 27.3\%. Glycolysis was slightly overestimated (median RPD: 7.4\%), with a median ARPD of 21.5\%. Reliability was excellent, with ICC values of 0.97 for volume and 1.00 for glycolysis.

For head and neck cancer, the model achieved a median Dice score of 0.71. Lesion volume was underestimated with a median RPD of –30.1\%, while ARPD reached a median of 36.1\%. Glycolysis was more robustly captured, with a median RPD of –11.9\% and a median ARPD of 20.7\%. Both measures showed good reliability, with ICC values of 0.85 and 0.98, respectively.

For metastatic melanoma, our model achieved a median Dice score of 0.69. Similar to head and neck cancer, lesion volumes were systematically underestimated (median RPD: –31.5\%), with an ARPD of 43.1\%. Glycolysis demonstrated closer agreement, with a median RPD of –16.1\% and a median ARPD of 25.1\%. Reliability was excellent for both lesion volume (ICC=0.98) and glycolysis (ICC=1.00).

\subsection{Inter-observer Variability Between Two Radiation Oncologists}

Overall lesion detection agreement for metastatic melanoma lesions between observers yielded an F1 score of 0.70. On a per-patient basis, the mean F1 score was 0.79 $\pm$ 0.18 with a 95\% confidence interval of 0.68--0.90. Segmentation overlap between observers showed a median patient-wise Dice similarity coefficient of 0.53 and a median lesion-wise Dice coefficient of 0.63.

For quantitative PET metrics, agreement between observers was generally high despite systematic differences in lesion extent. For TTB, MRPD was $-23.7\%$ with an interquartile range (IQR) of $-35.6\%$ to $-20.1\%$, and the MARPD was $33.0\%$ (IQR: $22.5\%$--$64.8\%$). The intraclass correlation coefficient indicated strong agreement ($\text{ICC}=0.95$). For TLG, the MRPD was $-6.3\%$ (IQR: $-19.8\%$ to $10.9\%$) and the MARPD was $17.6\%$ (IQR: $10.0\%$--$42.5\%$), with an $\text{ICC}$ of 0.96. On a lesion-wise level, volume measurements showed an MRPD of $-23.1\%$ (IQR: $-36.8\%$ to $3.2\%$) and a MARPD of $31.7\%$ (IQR: $15.1\%$--$47.8\%$), with very high agreement ($\text{ICC}=0.98$). Lesion-wise glycolysis measurements demonstrated an MRPD of $-3.4\%$ (IQR: $-24.0\%$ to $29.1\%$) and a MARPD of $25.7\%$ (IQR: $9.6\%$--$41.0\%$), with near-perfect agreement ($\text{ICC}=1.00$).

\section{Discussion}

This work introduces GLOW-FDG, an open-source whole-body cancer lesion segmentation model for FDG-PET/CT that was trained on what is, to our knowledge, one of the most diverse and heterogeneous curated datasets reported to date (Table \ref{tab:datasets}) and validated on cohorts with diverse cancer phenotypes. Our results show that GLOW-FDG had stable performance and consistently outperformed existing publicly available models across breast cancer, non- and oligometastatic lung cancer, head and neck cancer, and metastatic melanoma, yielding higher patient- and lesion-wise F1 scores by combining high sensitivity with a substantially lower false-positive rate. This balance is clinically relevant, since false-positive detections can lead to erroneous upstaging, for instance shifting patients from oligometastatic to polymetastatic disease, which in turn affects treatment decisions and may deny access to potentially curative approaches such as surgery or metastasis-directed radiotherapy. By reducing false positives without compromising recall, GLOW-FDG can improve staging reliability and strengthen trust in AI-assisted decision-making.

We benchmark our model against wide range of publicly available approaches: the multitracer model (AutoPET DKFZ), trained to segment both FDG- and PSMA-PET/CT images -- winner of the AutoPET III challenge; an FDG-PET/CT model developed by IKIM with dedicated preprocessing (AutoPET IKIM), which was part of a submission that achieved second place at AutoPET III; and a promising FDG-PET–only model (onlyPET) capable of segmenting images without requiring CT. Comparisons to benchmark models highlight the strengths and trade-offs of different approaches, the AutoPET DKFZ and AutoPET IKIM models achieved high recall but suffered from a high number of false-positive detections, which reduced precision and inflated volumetric biomarkers such as total tumor burden and total lesion glycolysis. The onlyPET performed well in head and neck cancer but had problems with metastatic melanoma predictions. Postprocessing was also applied for the AutoPET IKIM model based on SUV thresholding to mitigate erroneous predictions. While such heuristics can suppress noise, they introduce systematic biases and risk discarding true lesions with atypical uptake, an issue also noted in the broader literature on threshold-based PET segmentation \cite{Faro2026DLTumorBurdenMelanoma}. In contrast, GLOW-FDG avoided reliance on fixed thresholds, achieving strong lesion delineation and volumetric reproducibility across datasets without compromising generalizability. Importantly, its glycolysis quantification remained robust, even in cancer types where volumetric underestimation occurred, supporting its potential role in deriving prognostic imaging biomarkers. 

When compared to the performance of an expert radiation oncologists, the agreement was within the range of variability observed between clinicians. In particular, lesion detection showed comparable F1 scores, and segmentation overlap measured by the Dice similarity coefficient was similar to the level of agreement observed between the two observers. Quantitative PET-derived biomarkers, including total tumor burden and total lesion glycolysis, also demonstrated high concordance, with intraclass correlation coefficients comparable to those obtained between clinicians. These findings suggest that the segmentation and quantification performance of GLOW-FDG approaches the level of consistency typically observed between expert readers, supporting its potential utility for automated analysis of whole-body FDG-PET/CT.

Recent advances in automatic FDG-PET/CT segmentation have primarily been driven by training on AutoPET data \cite{Gatidis2022FDGPETCT}, and sometimes supplemented with HECKTOR data \cite{Oreiller2022HECKTORChallenge}, with only rare inclusion of additional external datasets for training or validation. Moreover, most studies focus primarily on spatial and volumetric agreement between ground truth and AI segmentations \cite{rokuss2024fdgpsmahitchhikersguide, kalisch2024autopetiiichallengeincorporating, Leung241979, He2023, 10.3389/fonc.2023.1235461}, with a strong emphasis on metrics such as the Dice similarity coefficient as also highlighted in \cite{ShiyamSundar2024}, while clinically relevant biomarkers such as TTB or TLG remain underexplored. Another observed limitation is that many publications present new model developments without releasing the trained weights \cite{Leung241979, He2023, 10.3389/fonc.2023.1235461}, making independent validation impossible and limiting the extent to which the field can benefit from these findings.

GLOW-FDG may serve as an out-of-the-box research and workflow-support tool and may help bridge the gap between methodological development and practical clinical application. In particular, automated pre-segmentation of FDG-avid lesions could reduce the time required for manual delineation and improve consistency during physician review, while also acting as a verification tool for potentially missed lesions. In addition, automated extraction of quantitative PET biomarkers such as TTB and TLG may facilitate large-scale imaging studies and clinical research where manual analysis would be impractical. Such automated analyses could also support investigations of longitudinal disease burden across serial PET/CT scans. GLOW-FDG is released as an open-source model with publicly available weights and code: https://github.com/MIC-DKFZ/GLOW-FDG

A key limitation of this study is that ultra–low-dose FDG-PET acquisitions were not included in either the training or validation datasets; therefore, the performance of GLOW-FDG under substantially reduced count statistics remains to be investigated. Furthermore, the model is trained to segment lesions that are visually detectable on FDG-PET, and lesions without clear PET visibility may not be reliably detected or segmented. The current version of GLOW-FDG was trained on FDG-PET/CT only, and while FDG remains the most widely used, incorporating other tracers such as PSMA, $^{68}Ga$, or DOTATATE would be a valuable future extension. However, the number of publicly available non-FDG datasets is considerably lower compared to FDG datasets, which hinders further progress in this direction due to current limited data availability. 

\section{Materials and Methods}
\subsection{Model architecture}
The lesion segmentation model was developed within the nnU-Net framework \cite{isensee2021nnu}. We adopted the ResEncL U-Net architecture preset \cite{isensee2024nnu}, which extends the standard encoder--decoder design with residual blocks and increased feature capacity. For multitask learning, an additional prediction head was added to segment organs with known high physiologic uptake of FDG (spleen, kidneys, liver, urinary bladder, lung, brain, heart, stomach, prostate, parotid and submandibular glands). The dual-headed design allows the network to jointly optimize lesion and organ segmentation, thereby reducing false positives and improving generalization. Both prediction heads were optimized using softmax cross-entropy combined with Dice loss at equal weights.

\subsection{Training Datasets}
The training data (Table \ref{tab:datasets}) represent a diverse set of internal institutional cohorts and publicly available MICCAI and TCIA datasets \cite{Gatidis2022FDGPETCT, Oreiller2022HECKTORChallenge, DeepPSMA2025, kinahan2019acrin6685, vallieres2017headneckpetct, bakr2017nsclc, PLESS20151049, Vallieres2015SoftTissueSarcoma, 10.1158/1078-0432.CCR-20-0020}. From the AutoPET challenge dataset \cite{Gatidis2022FDGPETCT}, we included 852 FDG-PET/CT scans in DICOM format of patients with lymphoma,  melanoma, lung cancer, and healthy controls. The HECKTOR challenge dataset \cite{Oreiller2022HECKTORChallenge} contributed 182 NIfTI PET/CT scans of head and neck cancer patients, and DEEP-PSMA challenge \cite{DeepPSMA2025} provided 75 NIfTI whole-body PET/CT scans of metastatic prostate cancer. Additional head and neck cancer cases came from TCIA datasets: ACRIN-HNSCC \cite{kinahan2019acrin6685} (65 scans, DICOM) and HN-PET-CT \cite{vallieres2017headneckpetct} (34 scans, DICOM). For lung cancer, our cohort used 67 scans from the TCIA-available NSCLC-RadGen dataset \cite{bakr2017nsclc} and 61 scans from the SAKK cohort \cite{PLESS20151049} available at our institution (both DICOM datasets). Another 14 DICOM FDG-PET/CT scans of soft tissue sarcoma were taken from the TCIA STS dataset \cite{Vallieres2015SoftTissueSarcoma} and institutional cohort from USZ -- SINERGIA \cite{10.1158/1078-0432.CCR-20-0020}, contributed 213 DICOM cases of metastatic melanoma and patients with complete recurrence after metastatic melanoma. In total, the training set consisted of 1563 FDG-PET/CT scans.

\subsection{Validation Datasets}
A total of five datasets (Table \ref{tab:datasets}) were used for validation. Two TCIA available datasets: QIN-Breast \cite{li2016qinbreast} with breast cancer (36 scans, DICOM) and ACRIN-NSCLC \cite{Kinahan2019HNSCC} with nonmetastatic lung cancer (35 scans, DICOM), and three private datasets available at USZ, including CHESS dataset \cite{GUCKENBERGER2025108553} with oligometastatic lung cancer (31 scans, DICOM), HECKTOR validation dataset from USZ \cite{Oreiller2022HECKTORChallenge} with head and neck cancer (63 scans, DICOM), and SINERGIA dataset \cite{10.1158/1078-0432.CCR-20-0020} with whole-body metastatic melanoma (20 scans, DICOM). Among these, only the SINERGIA dataset was split into training and validation subsets for model development; all other datasets were completely unseen during training and served as independent validation cohorts. Breast cancer \cite{li2016qinbreast} and nonmetastatic lung cancer \cite{Kinahan2019HNSCC} datasets from TCIA were initially segmented by a pretrained model for 3D semantic image segmentation of the FDG-avid lesions from PT/CT scans as a part of image segmentations produced by BAMF under the AIMI Annotations initiative \cite{van_oss_2024_bamf}, upon segmentation quality validation, it was observed that predicted segmentations typically overestimated lesion size, therefore, these two datasets were re-evaluated by medical doctors and used only for detection evaluation.

\subsection{Imaging Data Preprocessing}
PET data conversion to body-weight–normalized Standardized Uptake Value ($SUV_{BW}$) was performed according to
vendor-specific approach \cite{fritsak2025technicalnotevendorspecificapproach}, using the software Z-Rad \cite{FRITSAK2025S3712}, when DICOM series were available. When only NIfTI files were provided, it was assumed that the data providers had already performed the correct $SUV_{BW}$ conversion, and images were discarded only when the intensity in the normal tissues deviated strongly from the expected values during visual inspection.

All imaging data (Table \ref{tab:datasets}) were manually reviewed, and each PET image was checked to ensure that segmentation masks matched the corresponding lesions (i.e., no misalignment due to patient movement or breathing and masks were not substantially larger or smaller than the PET-visible part of the cancer lesion), that no cancer lesions were missed, all segmented structures represented true lesions (and not, for example, a needle tract, tracer injection site, or part of the ureter), and all segmented lesions were PET-visible. Image series not satisfying these criteria were excluded.

\subsection{Model training}
Model development followed a two-stage strategy. First, we pretrained a foundation model on a large, diverse corpus of medical imaging datasets, spanning CT, PET, and MRI modalities. Inspired by the MultiTalent framework\cite{ulrich2023multitalent}, this stage encouraged the network to acquire modality-invariant features and a broad anatomical understanding. Separate segmentation heads were employed for each dataset, and training was performed for 4,000 epochs with a patch size of 192$\times$192$\times$192 voxels and batch size of 24. Sampling across datasets was balanced inversely to the square root of dataset size.\\

\noindent After pretraining, the model was fine-tuned on the FDG-PET/CT datasets described above. Training lasted 1,500 epochs with stochastic gradient descent (Nesterov momentum 0.99, initial learning rate $10^{-2}$ with polynomial decay, batch size 3). Data augmentation included random rotations ($\leq 5^{\circ}$), elastic deformations, intensity scaling, and a misalignment augmentation\cite{kovacs2023addressing} to address patient motion and misregistration between PET and CT.

\subsection{Benchmark Models}
To compare our model against publicly available approaches, we benchmarked three external methods. First, we evaluated the winning model of the AutoPET III challenge developed in German Cancer Research Center (AutoPET DKFZ) \cite{rokuss2024fdgpsmahitchhikersguide}. This method is based on the nnU-Net framework using the ResEncL architecture and was developed for multitracer PET/CT lesion segmentation. It combines PET and CT as input and incorporates several task-specific strategies, including PET/CT misalignment augmentation, multimodal pretraining on CT/MR/PET datasets, and auxiliary organ supervision through a multi-task learning setup to better distinguish physiologic from pathologic uptake.

Second, we evaluated the FDG-PET/CT submodel of the second-place AutoPET III solution developed in Institute for AI in Medicine (AutoPET IKIM) \cite{kalisch2024autopetiiichallengeincorporating}, this approach also incorporates anatomical priors derived from TotalSegmentator-based organ labels during training. Their pipeline first classifies the tracer type and then applies a tracer-specific nnU-Net ensemble. For the FDG setting used here, the corresponding FDG-PET/CT model was combined with the authors’ dedicated post-processing, including SUV-based thresholding, as described in the original publication. 

Third, we included a publicly available FDG-PET-only lesion segmentation model (onlyPET) \cite{salimi_lesion_segmentation}. This model was trained within the nnU-Net framework for PET-only lesion segmentation and was selected as a unimodal comparator to assess performance without CT information. Inference was performed using the authors’ public implementation and recommended settings.

\subsection{Evaluation}
To establish consistent evaluation, when multiple AI segmentations corresponded to a single ground truth (GT) mask, they were combined and treated as one; likewise, when multiple GT masks corresponded to a single AI segmentation, these GT masks were also combined. This ensured a one-to-one correspondence between overlapping AI and GT cancer lesion masks.

Lesion detection performance was assessed for all studied models at the patient level across datasets using precision, recall, and F1 score with mean values, standard deviations (SD), and 95\% confidence intervals (95\% CI). To estimate 95\% CI, 10,000 bootstrap resamples with replacement were used. Additionally, performance was evaluated at the lesion level across each dataset to measure overall precision, recall, and F1 score.

Two-sided Mann-Whitney U tests with Benjamini-Hochberg correction were used to perform group-wise analysis to assess differences between the mean lesion $SUV_{BW}$, maximum lesion $SUV_{BW}$, lesion volume, and lesion glycolysis distributions across true positive (TP), false negative (FN), and false positive (FP) lesion groups for the GLOW-FDG model, where lesion glycolysis (in g) was defined as lesion volume (in mL) multiplied by mean $SUV_{BW}$ (in g/mL) in that volume. A p-value $<$ 0.05 was considered statistically significant.

Segmentation quality across all models was assessed at the patient level using the patient-wise Dice similarity score, total tumor burden (TTB), and total lesion glycolysis (TLG). Segmentation performance of GLOW-FDG was further evaluated at the lesion level using the lesion-wise Dice similarity score, lesion volume, and lesion glycolysis. For lesion-wise analyses, only TP lesions were included. To assess the accuracy of both patient- and lesion-wise measurements, the relative percentage difference (RPD), absolute relative percentage difference (ARPD), and intraclass correlation coefficient (3,1) (ICC) were calculated, collectively quantifying agreement and variability.

\subsection{Inter-observer Variability Between Two Radiation Oncologists}
To evaluate inter-observer variability in lesion detection and segmentation between clinicians, we computed the F1 score, Dice similarity coefficient, and ICC. Two certified radiation oncologists (MD1 and MD2) independently performed manual lesion delineations for 10 randomly selected whole-body metastatic melanoma cases from the validation dataset. To better contextualize the model’s performance, MD1 primarily focused on the FDG-PET signal (a model-like segmentation approach), whereas MD2 primarily relied on CT (a clinically typical segmentation style). 

\clearpage
\section{Appendix}

\subsection*{Pretraining dataset composition}
The pretraining stage combined a wide range of 3D medical imaging datasets, including CT-based organ segmentation collections (e.g. TotalSegmentator\cite{wasserthal2023totalsegmentator}), PET datasets with lesion or organ labels, and MRI datasets for anatomical segmentation tasks. This strategy ensured exposure to heterogeneous imaging modalities and anatomical regions. Each dataset was resampled to 1\,mm isotropic voxel spacing and intensity-normalized according to modality-specific schemes (HU windowing for CT, SUV normalization for PET, and Z-scoring for MRI).  

\subsection*{Pretraining configuration}
Pretraining was performed for 4{,}000 epochs on patches of size 192$\times$192$\times$192 voxels with a batch size of 24. Training used stochastic gradient descent with Nesterov momentum (0.99) and an initial learning rate of $10^{-2}$. Separate output heads were trained for each dataset, with sampling across datasets performed inversely proportional to the square root of dataset size. The objective combined Dice and cross-entropy loss with equal weights.  

\subsection*{Misalignment augmentation}
To address patient motion and misregistration between PET and CT, we introduced misalignment augmentation\cite{kovacs2023addressing}. For each patch, random rigid transformations were applied independently to the PET and CT channels, consisting of rotations up to $\pm 5^{\circ}$ and translations up to two voxels in the x and y directions. This augmentation was observed to improve robustness, particularly for small or punctate lesions where alignment noise otherwise degraded detection sensitivity.  

\subsection*{Organ supervision}
Organs selected for supervision were based on frequent physiologic FDG uptake and potential to confound lesion detection. TotalSegmentator predictions were generated for spleen, kidneys, liver, bladder, lung, brain, heart, stomach, prostate, and major salivary glands. These were used as auxiliary labels for the additional segmentation head. Losses from organ and lesion heads were averaged during optimization.  

\clearpage
\section{Declarations}

\noindent
 \textbf{Funding}: The conduct of this analysis was supported by Comprehensive Cancer Center Zurich (C3Z Precision Oncology Funding Program) OMD-ZH project. The ETOP CHESS study was sponsored and coordinated by the ETOP IBCSG Partners Foundation and conducted with support from AstraZeneca, as well as the foundation's own funds, as co-autor from ETOP IBCSG, with adress ETOP IBCSG Partners Foundation, Coordination Center, Bern, Switzerland: Rolf Stahel\\

\noindent
 \textbf{Conflicts of interest/Competing interests}: The Department of Radiation Oncology, University Hospital Zurich has research and teaching agreements with Siemens Healthineers (outside the scope of this work);  Prof. Dr. med. Martin Hüllner has received grants and speaker’s fees from GE HealthCare. Prof. Dr. med. Isabelle Opitz has no conflicts of interest. The following could be perceived as such: Roche (Institutional Grant), AstraZeneca (Advisory Board, Speakers Fee), MSD (Advisory Board and Interview), BMS (Advisory Board), Medtronic (Institutional Grant and Advisory Board), Intuitive (Proctorship and Speakers Fee), Sanofi (Speakers Fee), Regeneron (Advisory Board), XVIVO (Institutional Grant), Siemens (Speakers Fee), Astellas (Speakers Fee), Lilly (Speakers Fee). IO is IASLC Board Director, Member of the Thoracic Clinical Practice Standards Committee and the Thoracic Education Committee of AATS, iMig Board Member, SGT Board Member, JTCVS Associate Editor, and JAMA Surgery Editorial Board Member. She is in the Stiftungsrat Schulthessklinik and an Advisory Board Member at Med Uni Wien for Comprehensive Center for Chest Diseases (CCCD). Institutional roles: chair of the Lung Cancer Centre, member of the Robotic Board and Transplant Centre.\\

\noindent
 \textbf{Data availability statement}: AutoPET, HECKTOR, DEEP-PSMA, ACRIN-HNSCC, HN-PET-CT, NSCLC-RadGen, STS, QIN-Breast, and ACRIN-NSCLC are publically avaliable datasets. SINERGIA available upon reasonable request to the corresponding author. SAKK, HECKTOR-USZ, and CHESS cannot be shared. \\

\noindent
 \textbf{Acknowledgements}: We would like to acknowledge all participating sites that contributed to the creation of the validation datasets, as well as the sponsors who supported their development.\\

\noindent
 \textbf{Ethics approvals for data that are not publicly available}:
SINERGIA: Written informed consent was obtained from all patients, and the study was approved by the local ethics committee (Kantonale Ethikkommission Zürich, approval number 2019-01012) in accordance with “Good Clinical Practice” (GCP) guidelines and the Declaration of Helsinki; 
CHESS: The CHESS study protocol was approved by the local ethics committees and health authorities in all countries, and all patients gave written informed consent prior to enrolment. The study was registered on ClinicalTrials.gov (NCT03965468), and in the European Medicines Agency’s European Clinical Trials database, EU CT number 2024-511134-12. 
SAKK: Ethics amendments were received from all contributed Swiss Kanton ethics committees and informed consent was obtained from all individual participants (EKNZ PB\_2016-01071, KEK ZH PB\_2016-00412, KEK Bern PB\_2016-01072, CER-VD PB\_2016-01078, CCER PB\_2016-01073, EKOS PB\_2016-01075, Comitato Etico Cantonale Bellinzona PB\_2016-01077).

\clearpage
\newpage
\section{Tables and Figures}

\begin{figure}[ht]
    \centering
    \includegraphics[width=\textwidth]{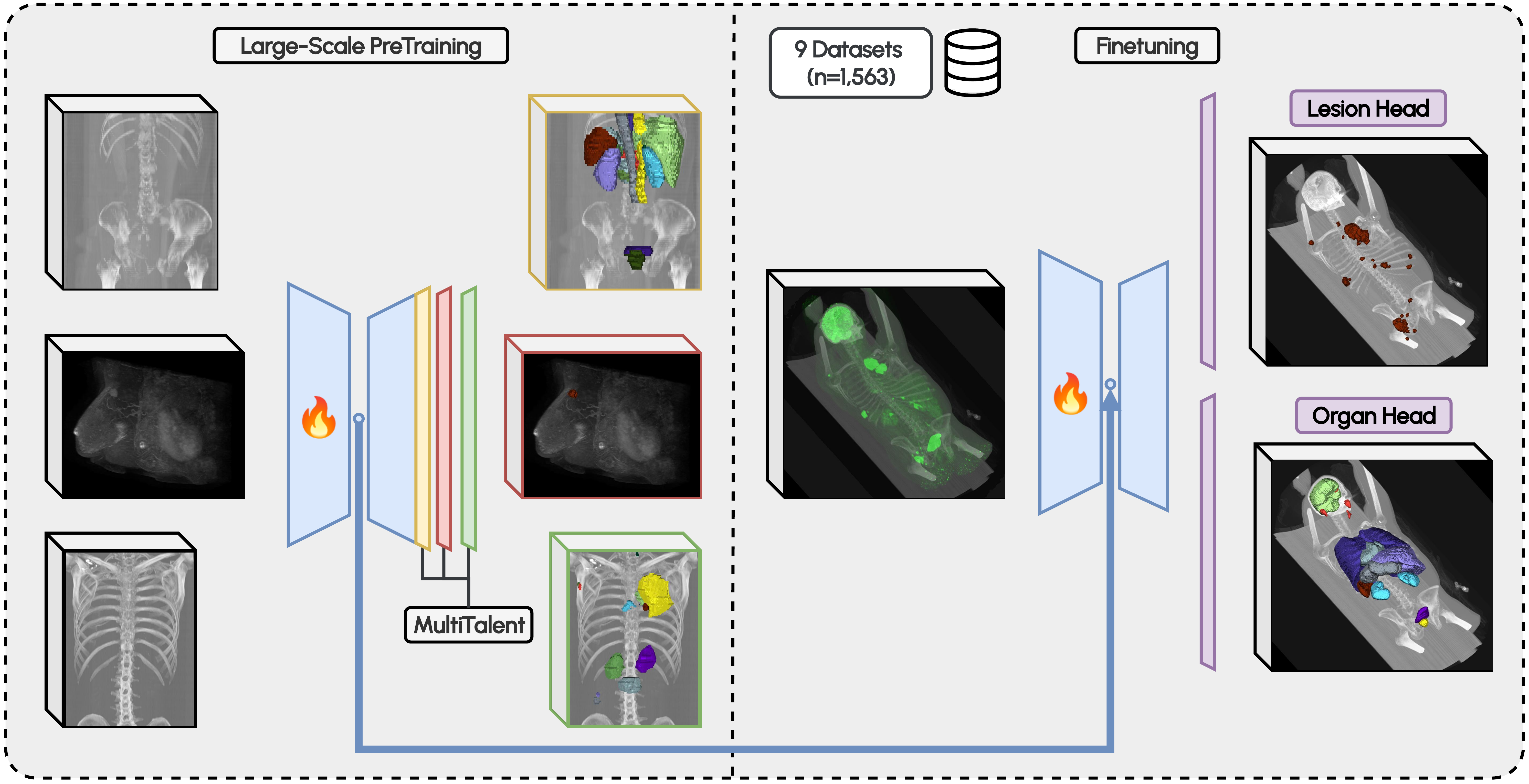}
    \caption{Overview of the GLOW-FDG training pipeline. The model first undergoes large-scale pretraining using a MultiTalent strategy across heterogeneous medical imaging tasks, then is finetuned on 1,563 FDG-PET/CT scans from 9 datasets. The finetuned network uses a dual-head design with one output for cancer lesion segmentation and a second auxiliary output for organ segmentation, helping distinguish pathological uptake from physiological FDG uptake.}
    \label{fig:overview}
\end{figure}

\begin{figure}[ht]
    \centering
    \includegraphics[width=\textwidth]{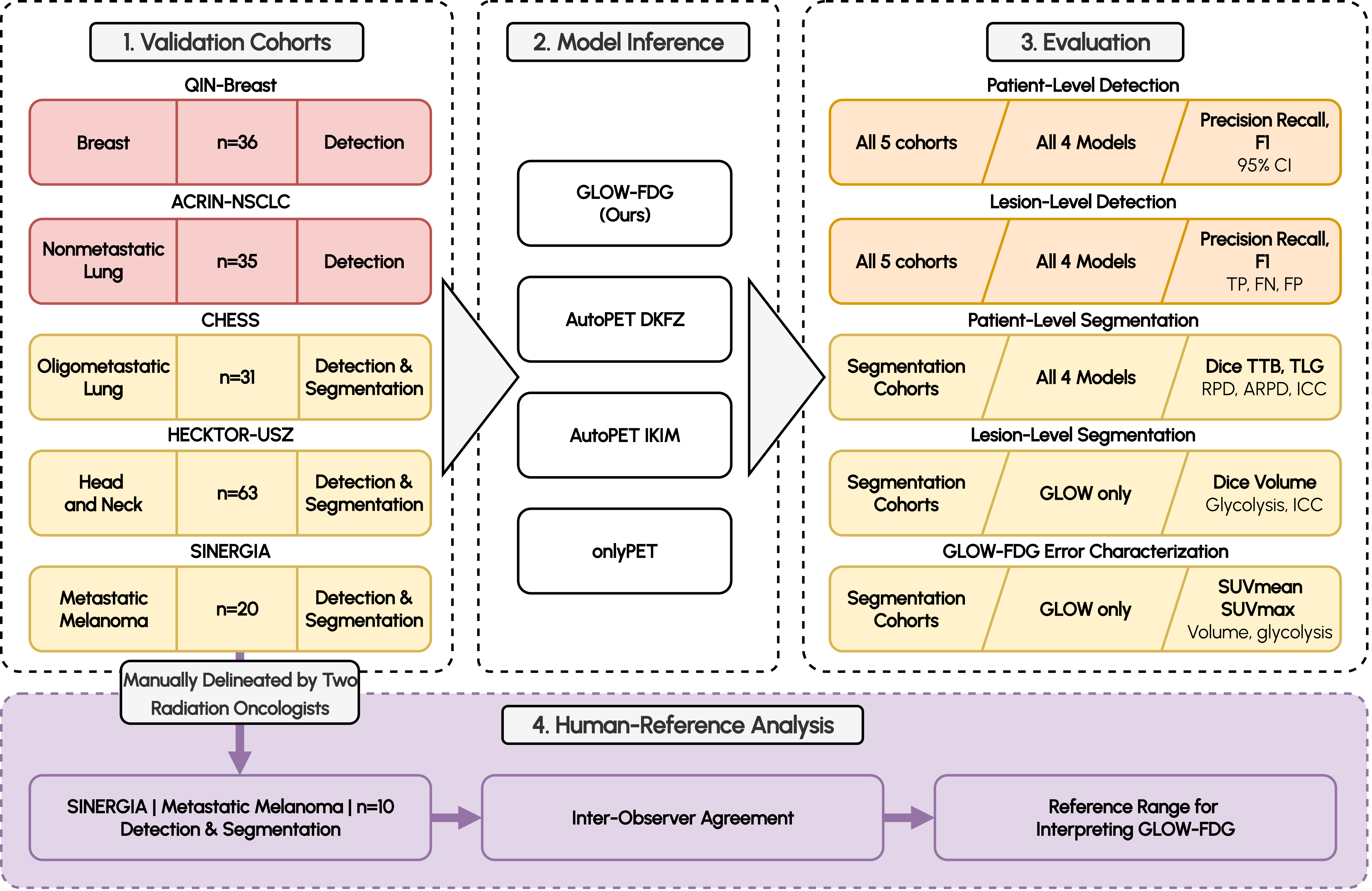}
    \caption{Overview of validation, benchmarking, and analysis workflow. Five independent FDG-PET/CT validation cohorts were used, covering breast cancer, nonmetastatic lung cancer, oligometastatic lung cancer, head and neck cancer, and metastatic melanoma. GLOW-FDG was compared with AutoPET DKFZ, AutoPET IKIM, and onlyPET for patient-level and lesion-level detection, while segmentation performance was assessed in the cohorts with manual lesion delineations. Additional analyses characterized GLOW-FDG errors and compared model performance with inter-observer variability from two radiation oncologists on a metastatic melanoma subset.}
    \label{fig:overview}
\end{figure}

\begin{table}[ht]
\centering
\caption{FDG-PET/CT Dataset Overview}
\resizebox{\textwidth}{!}{%
\begin{tabular}{l c c c c p{12cm}} 
\hline
\textbf{Dataset} & \textbf{Usage} & \makecell{\textbf{Cancer} \\ \textbf{type}} & \textbf{\# of Images} & \textbf{Data Type} & \textbf{Short Description} \\
\hline
AutoPET \cite{Gatidis2022FDGPETCT} & Training    & \makecell{Lymphoma \\ Melanoma\\ Lung Cancer} & 852 & DICOM & Whole body FDG-PET/CT data sets of patients with lymphoma, melanoma and non small cell lung cancer (NSCLC) as well as data sets without PET-positive malignant lesions (negative controls) examined between 2014 and 2018 at the University Hospital Tübingen. All examinations were acquired on a single, state-of-the-art PET/CT scanner (Siemens Biograph mCT). The imaging protocol consists of a diagnostic CT scan with intravenous contrast enhancement in most cases.\\
HECKTOR \cite{Oreiller2022HECKTORChallenge} & Training  & \makecell{Head and Neck \\ Cancer} & 182 & NIfTI & Patients with histologically confirmed oropharyngeal head and neck cancer who underwent radiotherapy and/or chemotherapy treatment planning were included. The dataset is based on FDG-PET and low-dose, non-contrast-enhanced CT images acquired using combined PET/CT scanners, We took data from CHUP, CHUM, and MDA.\\
DEEP-PSMA \cite{DeepPSMA2025} & Training & \makecell{Metastatic \\ Prostate Cancer} & 75 & NIfTI &   FDG-PET/CT was segmented using a liver-based threshold (mean SUV + 2*SD). Malignant vs. physiological uptake was manually annotated by an expert nuclear medicine physician ($\geq$5 years specialization). TTB labels were matched to PET resolution. Data were acquired mainly on GE Discovery 710/690, Siemens Biograph, and Siemens Vision 600 PET/CT scanners. Whole-body PET with low-dose CT was used for attenuation correction and localization, and PET was reconstructed with standard corrections and EANM EARL-compliant resolution recovery.\\
ACRIN-HNSCC \cite{kinahan2019acrin6685} & Training  & \makecell{Head and Neck \\ Cancer} & 65 & DICOM &  People with newly diagnosed head and neck squamous cell carcinoma being considered for surgical resection, with at least one side of the neck planned for dissection clinically N0, and at risk for occult metastasis (when risk based on clinical data is felt to be greater than 30\%). \\
HN-PET-CT \cite{vallieres2017headneckpetct}& Training & \makecell{Head and Neck \\ Cancer} & 34 & DICOM &  This collection contains FDG-PET/CT and radiotherapy planning CT imaging data of 298  patients from four different institutions in Québec with histologically proven head-and-neck cancer (H\&N) All patients had pre-treatment FDG-PET/CT scans between April 2006 and November 2014, and within a median of 18 days (range: 6-66) before treatment. This dataset is a part of HECKTOR challenge.\\
NSCLC-RadGen \cite{bakr2017nsclc}& Training   & \makecell{Metastatic and \\ Nonmetastatic \\ Lung Cancer} & 67 & DICOM &  Radiogenomic dataset from a Non-Small Cell Lung Cancer cohort. The dataset comprises Computed Tomography (CT), Positron Emission Tomography (PET)/CT images, semantic annotations of the tumors as observed on the medical images using a controlled vocabulary, segmentation maps of tumors in the CT scans, and quantitative values obtained from the PET/CT scans.\\
STS \cite{Vallieres2015SoftTissueSarcoma}& Training & \makecell{Soft Tissue \\ Sarcoma} & 14 & DICOM & Collection contains FDG-PET/CT imaging data with histologically proven soft-tissue sarcomas of the extremities; all patients had pre-treatment FDG-PET/CT scans between November 2004 and November 2011 \\
SAKK \cite{PLESS20151049} & Training & \makecell{Nonmetastatic \\ Lung Cancer} & 61 & DICOM &  The non-metastatic cohort consists of 34 FDG-PET/CT images of patients with pathologically proven, locally advanced T1–3N2M0, stage IIIA/N2 nonsmall-cell lung cancer, according to the sixth edition of the TNM classification. Staging was done by PET-CT and brain MRI. \\
SINERGIA \cite{10.1158/1078-0432.CCR-20-0020} & \makecell{Training and\\ Validation} & \makecell{Metastatic \\ Melanoma} & \makecell{Training: 213\\ Validation: 20} & DICOM & Single-institution deeply characterized metastatic melanoma cohort either single checkpoint inhibition (anti–PD-1) or dual checkpoint inhibition (anti–PD-1/anti–CTLA-4) treated between 2013 and 2019. Some patients show no indication of cancer on PET/CT images during the follow-up period at 3 and/or 6 months.\\
HECKTOR-USZ \cite{Oreiller2022HECKTORChallenge}& Validation & \makecell{Head and Neck \\ Cancer} & 63 & DICOM & Data description is the same as for the HECKTOR. This cohort is part of the HECKTOR validation and is not available online, thus used exclusively for the model validation. \\
CHESS \cite{GUCKENBERGER2025108553}& Validation & \makecell{Oligometastatic \\ Lung Cancer} & 31 & DICOM & The oligometastatic dataset comprises of FDG-PET/CT baseline scans from patients with synchronous oligometastatic non-small cell lung cancer (NSCLC). These patients were enrolled in the prospective, single-arm, international, multicenter phase II ETOP CHESS trial conducted between November 2019 and July 2022. \\
QIN-Breast \cite{li2016qinbreast}& Validation & \makecell{Breast \\ cancer} & 36 & DICOM & This dataset consists of FDG-PET/CT images obtained for treatment response assessment in breast cancer patients receiving neoadjuvant therapy. Imaging was performed on a GE Discovery STE scanner, using low-dose CT for attenuation correction and an FDG dose of approximately 370 MBq, adjusted according to patient body weight. \\
ACRIN-NSCLC \cite{Kinahan2019HNSCC}& Validation & \makecell{Nonmetastatic \\ Lung Cancer} & 35 & DICOM & Patients underwent concurrent platinum-based chemoradiotherapy (without surgery), with optional post-radiotherapy consolidation chemotherapy. Whole-body FDG-PET was performed at baseline on ACRIN-qualified scanners and repeated ~14 weeks after radiotherapy ($\geq$4 weeks after adjuvant chemotherapy), within 12–16 weeks of treatment completion, using the same scanner as baseline. \\
\hline
\end{tabular}}
\label{tab:datasets}
\end{table}

\begin{table}[htbp]
\centering
\caption{Lesion-wise precision, recall, and F1 score}
\begin{tabular}{clcccc}
\toprule
\multirow{2}{*}{Cancer Type} & \multirow{2}{*}{Metric} & \multicolumn{4}{c}{Models} \\
\cmidrule(lr){3-6}
 & & GLOW-FDG & AutoPET DKFZ & AutoPET IKIM & onlyPET \\
\midrule

\multirow{3}{*}{\shortstack{Breast \\ Cancer}} 
 & Precision  & $\bm{0.98}$ & 0.59 & 0.43 & 0.72 \\
 & Recall & 0.94 & 0.96 & $\bm{1.0}$ & $\bm{1.0}$ \\
 & F1  & $\bm{0.96}$ & 0.73 & 0.60 & 0.84 \\
\midrule

\multirow{3}{*}{\shortstack{Nonmetastatic \\ Lung Cancer}} 
 & Precision  & $\bm{0.97}$ & 0.76 & 0.74 & 0.76 \\
 & Recall & 0.98 & $\bm{1.0}$ & $\bm{1.0}$ & 0.90 \\
 & F1  & $\bm{0.97}$ & 0.87 & 0.85 & 0.83 \\
\midrule

\multirow{3}{*}{\shortstack{Oligometastatic \\ Lung Cancer}} 
 & Precision  & $\bm{0.83}$ & 0.70 & 0.63 & 0.69 \\
 & Recall & 0.86 & $\bm{0.89}$ & $\bm{0.89}$ & 0.79 \\
 & F1  & $\bm{0.85}$ & 0.79 & 0.74 & 0.73 \\
\midrule

\multirow{3}{*}{\shortstack{Head and Neck \\ Cancer}} 
 & Precision  & $\bm{0.90}$ & 0.52 & 0.54 & 0.80 \\
 & Recall & $\bm{0.87}$ & 0.44 & 0.66 & $\bm{0.87}$ \\
 & F1  & $\bm{0.89}$ & 0.48 & 0.59 & 0.83 \\
\midrule

\multirow{3}{*}{\shortstack{Metastatic \\ Melanoma}} 
 & Precision  & $\bm{0.84}$ & 0.71 & 0.61 & 0.65 \\
 & Recall & 0.80 & $\bm{0.84}$ & 0.77 & 0.62 \\
 & F1  & $\bm{0.82}$ & 0.77 & 0.68 & 0.64 \\
\bottomrule
\end{tabular}
\label{tab:lesion_detection}
\end{table}

\begin{table}[ht]
\centering
\caption{GLOW-FDG: lesion-wise segmentation results for volume and glycolysis}
\begin{tabular}{clcccccc}
\toprule
\multirow{2}{*}{\textbf{Cancer type}} & \multirow{2}{*}{} 
 & \multicolumn{2}{c}{\textbf{RPD [\%]}} 
 & \multicolumn{2}{c}{\textbf{ARPD [\%]}} 
 & \multirow{2}{*}{\textbf{ICC(3,1)}} \\
\cmidrule(lr){3-4} \cmidrule(lr){5-6}
 & & \textbf{Median} & \textbf{IQR} & \textbf{Median} & \textbf{IQR} & \\
\midrule
\addlinespace
\multirow{2}{*}{\makecell{Oligometastatic \\ Lung Cancer}} 
  & Volume     & 0.4   & -21.5 -- 30.5 & 27.3 & 13.6 -- 53.1 & 0.97 \\
  & Glycolysis & 7.4   & -6.7 -- 30.0  & 21.5 & 7.3 -- 42.1  & 1.00 \\
\addlinespace
\hline
\addlinespace
\multirow{2}{*}{\makecell{Head and Neck \\ Cancer}} 
  & Volume     & -30.1 & -46.8 -- -8.9 & 36.1 & 18.4 -- 48.7 & 0.85 \\
  & Glycolysis & -11.9 & -30.0 -- 2.8  & 20.7 & 8.8 -- 35.7  & 0.98 \\
\addlinespace
\hline
\addlinespace
\multirow{2}{*}{\makecell{Metastatic \\ Melanoma}} 
  & Volume     & -31.5 & -58.2 -- -4.9 & 43.1 & 21.1 -- 64.4 & 0.98 \\
  & Glycolysis & -16.1 & -41.3 -- 0.5  & 25.1 & 10.1 -- 50.7 & 1.00 \\
\addlinespace
\hline
\end{tabular}
\label{tab:glow_fdg_vol_lg}
\end{table}

\clearpage
\subsection{Supplements}

\begin{table}[htbp]
\centering
\caption{Patient-wise precision across models}
\begin{tabular}{clcccc}
\toprule
\multirow{2}{*}{Cancer Type} & \multirow{2}{*}{Precision} & \multicolumn{4}{c}{Models} \\
\cmidrule(lr){3-6}
 & & GLOW-FDG & AutoPET DKFZ & AutoPET IKIM & onlyPET \\
\midrule
\multirow{3}{*}{\shortstack{Breast \\ Cancer}}  
  & Mean  & 0.99 & 0.83 & 0.60 & 0.90 \\
  & SD    & 0.08 & 0.30 & 0.33 & 0.23 \\
  & 95\% CI & 0.96 -- 1.00 & 0.73 -- 0.92 & 0.50 -- 0.71 & 0.82 -- 0.97 \\
\midrule
\multirow{3}{*}{\shortstack{Nonmetastatic \\ Lung Cancer}}       
  & Mean  & 0.98 & 0.82 & 0.82 & 0.85 \\
  & SD    & 0.08 & 0.23 & 0.25 & 0.24 \\
  & 95\% CI & 0.95 -- 1.00 & 0.74 -- 0.90 & 0.73 -- 0.90 & 0.77 -- 0.92 \\
\midrule
\multirow{3}{*}{\shortstack{Oligometastatic \\ Lung Cancer}}        
  & Mean  & 0.88 & 0.73 & 0.67 & 0.75 \\
  & SD    & 0.17 & 0.22 & 0.22 & 0.23 \\
  & 95\% CI & 0.82 -- 0.94 & 0.65 -- 0.81 & 0.59 -- 0.75 & 0.67 -- 0.83 \\
\midrule
\multirow{3}{*}{\shortstack{Head and Neck \\ Cancer}}  
  & Mean  & 0.92 & 0.49 & 0.61 & 0.88 \\
  & SD    & 0.19 & 0.40 & 0.34 & 0.24 \\
  & 95\% CI & 0.87 -- 0.97 & 0.39 -- 0.59 & 0.52 -- 0.69 & 0.81 -- 0.93 \\
\midrule
\multirow{3}{*}{\shortstack{Metastatic \\ Melanoma}}      
  & Mean  & 0.83 & 0.68 & 0.61 & 0.57 \\
  & SD    & 0.24 & 0.32 & 0.38 & 0.35 \\
  & 95\% CI & 0.71 -- 0.92 & 0.54 -- 0.81 & 0.44 -- 0.77 & 0.42 -- 0.73 \\
\bottomrule
\end{tabular}
\label{tab:pat_prec}
\end{table}

\begin{table}[htbp]
\centering
\caption{Patient-wise recall across models}
\begin{tabular}{clcccc}
\toprule
\multirow{2}{*}{Cancer Type} & \multirow{2}{*}{Recall} & \multicolumn{4}{c}{Models} \\
\cmidrule(lr){3-6}
 & & GLOW-FDG & AutoPET DKFZ & AutoPET IKIM & onlyPET \\
\midrule
\multirow{3}{*}{\shortstack{Breast \\ Cancer}}
  & Mean  & 0.96 & 0.96 & 1.00 & 1.00 \\
  & SD    & 0.14 & 0.18 & 0.00 & 0.00 \\
  & 95\% CI & 0.90 -- 1.00 & 0.89 -- 1.00 & 1.00 -- 1.00 & 1.00 -- 1.00 \\
\midrule
\multirow{3}{*}{\shortstack{Nonmetastatic \\ Lung Cancer}}        
  & Mean  & 0.99 & 1.00 & 1.00 & 0.95 \\
  & SD    & 0.04 & 0.00 & 0.00 & 0.16 \\
  & 95\% CI & 0.98 -- 1.00 & 1.00 -- 1.00 & 1.00 -- 1.0 & 0.90 -- 1.0 \\
\midrule
\multirow{3}{*}{\shortstack{Oligometastatic \\ Lung Cancer}}      
  & Mean  & 0.88 & 0.91 & 0.91 & 0.80 \\
  & SD    & 0.19 & 0.15 & 0.15 & 0.21 \\
  & 95\% CI & 0.81 -- 0.94 & 0.85 -- 0.96 & 0.85 -- 0.96 & 0.73 -- 0.87 \\
\midrule
\multirow{3}{*}{\shortstack{Head and Neck \\ Cancer}} 
  & Mean  & 0.90 & 0.42 & 0.66 & 0.90 \\
  & SD    & 0.20 & 0.33 & 0.32 & 0.21 \\
  & 95\% CI & 0.85 -- 0.94 & 0.33 -- 0.50 & 0.58 -- 0.74 & 0.84 -- 0.95 \\
\midrule
\multirow{3}{*}{\shortstack{Metastatic \\ Melanoma}}    
  & Mean  & 0.85 & 0.82 & 0.77 & 0.61 \\
  & SD    & 0.24 & 0.30 & 0.35 & 0.38 \\
  & 95\% CI & 0.73 -- 0.94 & 0.68 -- 0.94 & 0.60 -- 0.90 & 0.44 -- 0.78 \\
\bottomrule
\end{tabular}
\label{tab:pat_recall}
\end{table}

\begin{table}[htbp]
\centering
\caption{Patient-wise F1 across models}
\begin{tabular}{clcccc}
\toprule
\multirow{2}{*}{Cancer Type} & \multirow{2}{*}{F1} & \multicolumn{4}{c}{Models} \\
\cmidrule(lr){3-6}
 & & GLOW-FDG & AutoPET DKFZ & AutoPET IKIM & onlyPET \\
\midrule
\multirow{3}{*}{\shortstack{Breast \\ Cancer}}
  & Mean  & 0.96 & 0.86 & 0.70 & 0.93 \\
  & SD    & 0.10 & 0.26 & 0.27 & 0.18 \\
  & 95\% CI & 0.93 -- 0.99 & 0.77 -- 0.94 & 0.61 -- 0.79 & 0.86 -- 0.98 \\
\midrule
\multirow{3}{*}{\shortstack{Nonmetastatic \\ Lung Cancer}}       
  & Mean  & 0.98 & 0.88 & 0.88 & 0.87 \\
  & SD    & 0.05 & 0.15 & 0.18 & 0.19 \\
  & 95\% CI & 0.96 -- 1.0 & 0.83 -- 0.93 & 0.81 -- 0.93 & 0.81 -- 0.93 \\
\midrule
\multirow{3}{*}{\shortstack{Oligometastatic \\ Lung Cancer}}
  & Mean  & 0.86 & 0.79 & 0.74 & 0.74 \\
  & SD    & 0.15 & 0.16 & 0.16 & 0.18 \\
  & 95\% CI & 0.80 -- 0.91 & 0.73 -- 0.84 & 0.69 -- 0.80 & 0.68 -- 0.80 \\
\midrule
\multirow{3}{*}{\shortstack{Head and Neck \\ Cancer}}
  & Mean  & 0.89 & 0.42 & 0.59 & 0.86 \\
  & SD    & 0.18 & 0.32 & 0.30 & 0.21 \\
  & 95\% CI & 0.85 -- 0.94 & 0.34 -- 0.50 & 0.52 -- 0.66 & 0.80 -- 0.91 \\
\midrule
\multirow{3}{*}{\shortstack{Metastatic \\ Melanoma}}    
  & Mean  & 0.83 & 0.72 & 0.63 & 0.56 \\
  & SD    & 0.22 & 0.29 & 0.35 & 0.33 \\
  & 95\% CI & 0.71 -- 0.91 & 0.58 -- 0.83 & 0.48 -- 0.78 & 0.41 -- 0.70 \\
\bottomrule
\end{tabular}
\label{tab:pat_f1}
\end{table}


\begin{table}[ht]
\centering
\caption{Oligometastatic Lung Cancer: TP, FN, and FP Distributions}
\begin{tabular}{llcccc}
\toprule
\textbf{Metric} & \textbf{} & \textbf{FP} & \textbf{FN} & \textbf{TP (AI)} & \textbf{TP (GT)} \\
\midrule
\multirow{2}{*}{Volume (mL)} 
  & Median & 0.19 & 1.09 & 9.53 & 8.93 \\
  & IQR    & 0.05 -- 0.54 & 0.76 -- 1.81 & 2.32 -- 36.08 & 3.73 -- 32.74 \\
\addlinespace
\midrule
\multirow{2}{*}{Lesion Glycolysis (g)} 
  & Median & 0.74 & 2.79 & 65.42 & 51.16 \\
  & IQR    & 0.24 -- 2.07 & 2.20 -- 4.29 & 10.63 -- 203.70 & 12.49 -- 183.73 \\
\addlinespace
\midrule
\multirow{2}{*}{Mean $SUV_{BW}$ (g/mL)} 
  & Median & 4.14 & 2.54 & 5.49 & 4.95 \\
  & IQR    & 3.66 -- 5.07 & 1.87 -- 3.56 & 4.14 -- 6.90 & 3.66 -- 6.77 \\
\addlinespace
\midrule
\multirow{2}{*}{Max. $SUV_{BW}$ (g/mL)} 
  & Median & 4.85 & 3.63 & 11.35 & 11.35 \\
  & IQR    & 4.12 -- 7.52 & 3.18 -- 4.93 & 7.68 -- 16.73 & 7.45 -- 16.73 \\
\bottomrule
\end{tabular}
\label{tab:tp_fn_fp_lung}
\end{table}

\begin{table}[ht]
\centering
\caption{Head and Neck Cancer: TP, FN, and FP Distributions}
\begin{tabular}{llcccc}
\toprule
\textbf{Metric} & \textbf{} & \textbf{FP} & \textbf{FN} & \textbf{TP (AI)} & \textbf{TP (GT)} \\
\midrule
\multirow{2}{*}{Volume (mL)} 
  & Median & 0.42 & 1.47 & 7.17 & 9.29 \\
  & IQR    & 0.25 -- 1.35 & 0.67 -- 2.70 & 3.29 -- 15.60 & 4.13 -- 19.82 \\
\addlinespace
\midrule
\multirow{2}{*}{Lesion Glycolysis (g)} 
  & Median & 1.64 & 3.18 & 47.94 & 49.51 \\
  & IQR    & 0.89 -- 4.91 & 1.23 -- 6.29 & 13.57 -- 93.71 & 18.39 -- 96.69 \\
\addlinespace
\midrule
\multirow{2}{*}{Mean $SUV_{BW}$ (g/mL)} 
  & Median & 3.89 & 2.11 & 6.42 & 4.90 \\
  & IQR    & 3.53 -- 5.43 & 1.78 -- 2.89 & 4.49 -- 8.25 & 3.73 -- 6.86 \\
\addlinespace
\midrule
\multirow{2}{*}{Max $SUV_{BW}$ (g/mL)} 
  & Median & 4.43 & 3.51 & 11.28 & 11.28 \\
  & IQR    & 3.61 -- 6.99 & 2.97 -- 3.85 & 7.70 -- 14.30 & 8.08 -- 14.30 \\
\bottomrule
\end{tabular}
\label{tab:tp_fn_fp_h_and_n}
\end{table}

\begin{table}[ht]
\centering
\caption{Metastatic Melanoma: TP, FN, and FP Distributions}
\begin{tabular}{llcccc}
\toprule
\textbf{Metric} & \textbf{} & \textbf{FP} & \textbf{FN} & \textbf{TP (AI)} & \textbf{TP (GT)} \\
\midrule
\multirow{2}{*}{Volume (mL)} 
  & Median & 0.20 & 1.27 & 1.97 & 2.69 \\
  & IQR    & 0.08 -- 0.59 & 0.75 -- 1.70 & 0.46 -- 6.79 & 1.08 -- 7.61 \\
\addlinespace
\midrule
\multirow{2}{*}{Lesion Glycolysis (g)} 
  & Median & 0.72 & 2.90 & 8.29 & 11.24 \\
  & IQR    & 0.20 -- 1.91 & 1.92 -- 5.16 & 1.86 -- 40.02 & 3.48 -- 38.36 \\
\addlinespace
\midrule
\multirow{2}{*}{Mean $SUV_{BW}$ (g/mL)} 
  & Median & 3.80 & 2.74 & 4.73 & 3.85 \\
  & IQR    & 3.10 -- 4.12 & 1.97 -- 3.59 & 3.98 -- 6.05 & 3.17 -- 5.11 \\
\addlinespace
\midrule
\multirow{2}{*}{Max $SUV_{BW}$ (g/mL)} 
  & Median & 4.21 & 4.41 & 7.56 & 7.75 \\
  & IQR    & 3.42 -- 5.78 & 3.28 -- 5.37 & 5.49 -- 11.81 & 5.75 -- 11.81 \\
\bottomrule
\end{tabular}
\label{tab:tp_fn_fp_melanoma}
\end{table}

\begin{table}[ht]
\centering
\caption{Oligometastatic Lung Cancer: Mann--Whitney U-Test p-values with Benjamini-Hochberg correction}
\begin{tabular}{llcccccc}
\toprule
\textbf{Metric} & \textbf{Comparison} & \textbf{p-value} \\
\midrule
\multirow{6}{*}{Volume (mL)} 
  & TP(GT) vs. TP(AI) & 0.978 \\
  & FP vs. FN         & $\bm{<0.001}$ \\
  & FN vs. TP(AI)     & $\bm{<0.001}$ \\
  & FN vs. TP(GT)     & $\bm{<0.001}$ \\
  & FP vs. TP(AI)     & $\bm{<0.001}$ \\
  & FP vs. TP(GT)     & $\bm{<0.001}$ \\
\addlinespace
\multirow{6}{*}{Lesion Glycolysis (g)} 
  & TP(GT) vs. TP(AI) & 0.838 \\
  & FP vs. FN         & $\bm{0.004}$ \\
  & FN vs. TP(AI)     & $\bm{<0.001}$ \\
  & FN vs. TP(GT)     & $\bm{<0.001}$ \\
  & FP vs. TP(AI)     & $\bm{<0.001}$ \\
  & FP vs. TP(GT)     & $\bm{<0.001}$ \\
\addlinespace
\multirow{6}{*}{Max $SUV_{BW}$ (g/mL)} 
  & TP(GT) vs. TP(AI) & 0.978 \\
  & FP vs. FN         & $\bm{0.046}$ \\
  & FN vs. TP(AI)     & $\bm{<0.001}$ \\
  & FN vs. TP(GT)     & $\bm{<0.001}$ \\
  & FP vs. TP(AI)     & $\bm{<0.001}$ \\
  & FP vs. TP(GT)     & $\bm{<0.001}$ \\
\addlinespace
\multirow{6}{*}{Mean $SUV_{BW}$ (g/mL)} 
  & TP(GT) vs. TP(AI) & 0.135 \\
  & FP vs. FN         & $\bm{0.003}$ \\
  & FN vs. TP(AI)     & $\bm{<0.001}$ \\
  & FN vs. TP(GT)     & $\bm{<0.001}$ \\
  & FP vs. TP(AI)     & $\bm{0.021}$ \\
  & FP vs. TP(GT)     & 0.268 \\
\bottomrule
\end{tabular}
\label{tab:p_lung}
\end{table}

\begin{table}[ht]
\centering
\caption{Head and Neck Cancer: Mann--Whitney U-Test p-values with Benjamini-Hochberg correction }
\begin{tabular}{llcccccc}
\toprule
\textbf{Metric} & \textbf{Comparison} & \textbf{p-value} \\
\midrule

\multirow{6}{*}{Volume (mL)} 
  & TP(GT) vs. TP(AI) & 0.051 \\
  & FP vs. FN         & 0.081 \\
  & FN vs. TP(AI)     & $\bm{<0.001}$ \\
  & FN vs. TP(GT)     & $\bm{<0.001}$ \\
  & FP vs. TP(AI)     & $\bm{<0.001}$ \\
  & FP vs. TP(GT)     & $\bm{<0.001}$ \\
\addlinespace
\multirow{6}{*}{Lesion Glycolysis (g)} 
  & TP(GT) vs. TP(AI) & 0.558 \\
  & FP vs. FN         & 0.640 \\
  & FN vs. TP(AI)     & $\bm{<0.001}$ \\
  & FN vs. TP(GT)     & $\bm{<0.001}$ \\
  & FP vs. TP(AI)     & $\bm{<0.001}$ \\
  & FP vs. TP(GT)     & $\bm{<0.001}$ \\
\addlinespace
\multirow{6}{*}{Max $SUV_{BW}$ (g/mL)} 
  & TP(GT) vs. TP(AI) & 0.992 \\
  & FP vs. FN         & $\bm{0.041}$ \\
  & FN vs. TP(AI)     & $\bm{<0.001}$ \\
  & FN vs. TP(GT)     & $\bm{<0.001}$ \\
  & FP vs. TP(AI)     & $\bm{0.002}$ \\
  & FP vs. TP(GT)     & $\bm{0.002}$ \\
\addlinespace
\multirow{6}{*}{Mean $SUV_{BW}$ (g/mL)} 
  & TP(GT) vs. TP(AI) & $\bm{<0.001}$ \\
  & FP vs. FN         & $\bm{0.002}$ \\
  & FN vs. TP(AI)     & $\bm{<0.001}$ \\
  & FN vs. TP(GT)     & $\bm{<0.001}$ \\
  & FP vs. TP(AI)     & $\bm{0.02}$ \\
  & FP vs. TP(GT)     & 0.28 \\
\bottomrule
\end{tabular}
\label{tab:p_h_and_n}
\end{table}

\begin{table}[ht]
\centering
\caption{Metastatic Melanoma: Mann--Whitney U-Test p-values with Benjamini-Hochberg correction}
\begin{tabular}{llcccccc}
\toprule
\textbf{Metric} & \textbf{Comparison} & \textbf{p-value} \\
\midrule
\multirow{6}{*}{Volume (mL)} 
  & TP(GT) vs. TP(AI) & $\bm{0.045}$ \\
  & FP vs. FN         & $\bm{0.002}$ \\
  & FN vs. TP(AI)     & 0.538 \\
  & FN vs. TP(GT)     & $\bm{0.005}$ \\
  & FP vs. TP(AI)     & $\bm{0.003}$ \\
  & FP vs. TP(GT)     & $\bm{<0.001}$ \\
\addlinespace
\multirow{6}{*}{Lesion Glycolysis (g)} 
  & TP(GT) vs. TP(AI) & 0.213 \\
  & FP vs. FN         & $\bm{0.014}$ \\
  & FN vs. TP(AI)     & 0.071 \\
  & FN vs. TP(GT)     & $\bm{0.001}$ \\
  & FP vs. TP(AI)     & $\bm{0.001}$ \\
  & FP vs. TP(GT)     & $\bm{<0.001}$ \\
\addlinespace
\multirow{6}{*}{Max $SUV_{BW}$ (g/mL)} 
  & TP(GT) vs. TP(AI) & 0.966 \\
  & FP vs. FN         & 0.978 \\
  & FN vs. TP(AI)     & $\bm{0.001}$ \\
  & FN vs. TP(GT)     & $\bm{0.001}$ \\
  & FP vs. TP(AI)     & $\bm{0.001}$ \\
  & FP vs. TP(GT)     & $\bm{0.001}$ \\
\addlinespace
\multirow{6}{*}{Mean $SUV_{BW}$ (g/mL)} 
  & TP(GT) vs. TP(AI) & $\bm{0.002}$ \\
  & FP vs. FN         & 0.067 \\
  & FN vs. TP(AI)     & $\bm{<0.001}$ \\
  & FN vs. TP(GT)     & $\bm{0.004}$ \\
  & FP vs. TP(AI)     & $\bm{0.002}$ \\
  & FP vs. TP(GT)     & 0.496 \\
\bottomrule
\end{tabular}
\label{tab:p_melanoma}
\end{table}

\begin{table}[htbp]
\centering
\caption{Patient-wise Dice across models}
\begin{tabular}{clcccc}
\toprule
\multirow{2}{*}{Cancer Type} & \multirow{2}{*}{Dice} & \multicolumn{4}{c}{Models} \\
\cmidrule(lr){3-6}
 & & GLOW-FDG & AutoPET DKFZ & AutoPET IKIM & onlyPET \\
\midrule
\multirow{3}{*}{\shortstack{Oligometastatic \\ Lung Cancer}}       
  & Mean  & 0.78 & 0.72 & 0.71 & 0.76 \\
  & SD & 0.11 & 0.11 & 0.12 & 0.13 \\
  & 95\% CI & 0.72 -- 0.79 & 0.69 -- 0.76 & 0.66 -- 0.74 & 0.67 -- 0.76 \\
\midrule
\multirow{3}{*}{\shortstack{Head and Neck \\ Cancer}} 
  & Mean  & 0.71 & 0.29 & 0.52 & 0.68 \\
  & SD & 0.17 & 0.27 & 0.27 & 0.16 \\
  & 95\% CI & 0.63 -- 0.71 & 0.24 -- 0.38 & 0.39 -- 0.52 & 0.62 -- 0.70 \\
\midrule
\multirow{3}{*}{\shortstack{Metastatic \\ Melanoma}}      
  & Mean  & 0.70 & 0.70 & 0.67 & 0.62 \\
  & SD & 0.24 & 0.26 & 0.31 & 0.28 \\
  & 95\% CI & 0.51 -- 0.72 & 0.48 -- 0.71 & 0.41 -- 0.68 & 0.37 -- 0.62 \\
\bottomrule
\end{tabular}
\label{tab:pat_Dice}
\end{table}

\begin{table}[htbp]
\centering
\caption{GLOW-FDG: Total tumor burden and total lesion glycolysis}
\begin{tabular}{clccccc}
\toprule
\multirow{2}{*}{\textbf{Cancer type}} & \multirow{2}{*}{} 
 & \multicolumn{2}{c}{\textbf{RPD [\%]}} 
 & \multicolumn{2}{c}{\textbf{ARPD [\%]}} 
 & \multirow{2}{*}{\textbf{ICC(3,1)}} \\
\cmidrule(lr){3-4} \cmidrule(lr){5-6}
 & & \textbf{Median} & \textbf{IQR} & \textbf{Median} & \textbf{IQR} & \\
\midrule
\multirow{2}{*}{\shortstack{Oligometastatic \\ Lung Cancer}} 
  & TTB & -0.5 & -12.9 -- 23.9 & 20.8 & 8.6 -- 31.1 & 0.96 \\
  & TLG & 4.5 & 0.4 -- 21.6 & 10.9 & 3.5 -- 27.1 & 0.99 \\
\midrule
\multirow{2}{*}{\shortstack{Head and Neck \\ Cancer}}  
  & TTB & -27.4 & -42.3 -- -9.2 & 29.5 & 15.4 -- 42.3 & 0.79 \\
  & TLG & -8.8 & -21.4 -- 4.3 & 16.8 & 8.4 -- 27.9 & 0.96 \\
\midrule
\multirow{2}{*}{\shortstack{Metastatic \\ Melanoma}}  
  & TTB & -22.3 & -53.7 -- -5.4 & 27.6 & 16.2 -- 57.7 & 0.98 \\
  & TLG & -9.9 & -35.9 -- -1.0 & 14.2 & 8.2 -- 43.8 & 1.00 \\
\bottomrule
\end{tabular}
\label{tab:pat_wise_GLOW}
\end{table}

\begin{table}[htbp]
\centering
\caption{AutoPET DKFZ: Total tumor burden and total lesion glycolysis}
\begin{tabular}{clccccc}
\toprule
\multirow{2}{*}{\textbf{Cancer type}} & \multirow{2}{*}{} 
 & \multicolumn{2}{c}{\textbf{RPD [\%]}} 
 & \multicolumn{2}{c}{\textbf{ARPD [\%]}} 
 & \multirow{2}{*}{\textbf{ICC(3,1)}} \\
\cmidrule(lr){3-4} \cmidrule(lr){5-6}
 & & \textbf{Median} & \textbf{IQR} & \textbf{Median} & \textbf{IQR} & \\
\midrule
\multirow{2}{*}{\shortstack{Oligometastatic \\ Lung Cancer}} 
  & TTB & 32.5 & 3.6 -- 69.1 & 33.9 & 10.2 -- 69.1 & 0.89 \\
  & TLG & 17.8 & 7.2 -- 42.7 & 22.7 & 8.7 -- 42.7 & 0.98 \\
\midrule
\multirow{2}{*}{\shortstack{Head and Neck \\ Cancer}} 
  & TTB & -60.7 & -94.6 -- -44.1 & 60.7 & 46.2 -- 94.6 & 0.61 \\
  & TLG & -56.3 & -95.4 -- -27.2 & 65.2 & 41.2 -- 97.9 & 0.70 \\
\midrule
\multirow{2}{*}{\shortstack{Metastatic \\ Melanoma}}  
  & TTB & -12.7 & -38.3 -- 8.1 & 30.3 & 8.9 -- 68.9 & 0.92 \\
  & TLG & -4.0 & -16.6 -- 13.1 & 14.2 & 6.3 -- 60.3 & 0.99 \\
\bottomrule
\end{tabular}
\label{tab:pat_wise_AutoPET_DKFZ}
\end{table}

\begin{table}[htbp]
\centering
\caption{AutoPET IKIM: Total tumor burden and total lesion glycolysis}
\begin{tabular}{clccccc}
\toprule
\multirow{2}{*}{\textbf{Cancer type}} & \multirow{2}{*}{} 
 & \multicolumn{2}{c}{\textbf{RPD [\%]}} 
 & \multicolumn{2}{c}{\textbf{ARPD [\%]}} 
 & \multirow{2}{*}{\textbf{ICC(3,1)}} \\
\cmidrule(lr){3-4} \cmidrule(lr){5-6}
 & & \textbf{Median} & \textbf{IQR} & \textbf{Median} & \textbf{IQR} & \\
\midrule
\multirow{2}{*}{\shortstack{Oligometastatic \\ Lung Cancer}} 
  & TTB & 45.4 & 21.5 -- 76.3 & 45.4 & 23.5 -- 76.3 & 0.79 \\
  & TLG & 24.8 & 13.4 -- 42.4 & 25.6 & 13.9 -- 42.4 & 0.96 \\
\midrule
\multirow{2}{*}{\shortstack{Head and Neck \\ Cancer}} 
  & TTB & -16.6 & -43.2 -- 11.8 & 25.5 & 13.3 -- 61.8 & 0.71 \\
  & TLG & -13.3 & -45.4 -- 16.8 & 32.0 & 16.2 -- 64.7 & 0.88 \\
\midrule
\multirow{2}{*}{\shortstack{Metastatic \\ Melanoma}}  
  & TTB & -7.6 & -27.1 -- 32.7 & 31.1 & 17.4 -- 90.8 & 0.95 \\
  & TLG & 0.6 & -16.4 -- 28.6 & 23.9 & 10.8 -- 88.0 & 0.99 \\
\bottomrule
\end{tabular}
\label{tab:pat_wise_AutoPET_IKIM}
\end{table}

\begin{table}[htbp]
\centering
\caption{onlyPET: Total tumor burden and total lesion glycolysis}
\begin{tabular}{clccccc}
\toprule
\multirow{2}{*}{\textbf{Cancer type}} & \multirow{2}{*}{} 
 & \multicolumn{2}{c}{\textbf{RPD [\%]}} 
 & \multicolumn{2}{c}{\textbf{ARPD [\%]}} 
 & \multirow{2}{*}{\textbf{ICC(3,1)}} \\
\cmidrule(lr){3-4} \cmidrule(lr){5-6}
 & & \textbf{Median} & \textbf{IQR} & \textbf{Median} & \textbf{IQR} & \\
\midrule
\multirow{2}{*}{\shortstack{Oligometastatic \\ Lung Cancer}} 
  & TTB & -9.7 & -16.1 -- 14.0 & 15.0 & 10.3 -- 25.2 & 0.92 \\
  & TLG & -0.3 & -6.5 -- 14.4 & 7.9 & 3.2 -- 21.0 & 0.97 \\
\midrule
\multirow{2}{*}{\shortstack{Head and Neck \\ Cancer}} 
  & TTB & -15.7 & -32.2 -- 1.6 & 22.0 & 13.0 -- 39.9 & 0.63 \\
  & TLG & -2.0 & -12.5 -- 14.3 & 13.5 & 4.9 -- 26.1 & 0.89 \\
\midrule
\multirow{2}{*}{\shortstack{Metastatic \\ Melanoma}}  
  & TTB & -28.1 & -42.0 -- 3.0 & 43.1 & 28.4 -- 75.4 & 0.90 \\
  & TLG & -14.2 & -27.4 -- 11.5 & 27.8 & 14.4 -- 62.6 & 0.97 \\
\bottomrule
\end{tabular}
\label{tab:pat_wise_onlyPET}
\end{table}

\begin{table}[ht]
    \centering
    \caption{GLOW-FDG: Confusion matrices for different validation datasets}

    \begin{subtable}[t]{0.45\textwidth}
        \centering
        \caption{Breast Cancer}
        \begin{tabular}{c|c c}
            & \shortstack{Predicted \\ Positive} & \shortstack{Predicted \\ Negative} \\
            \hline
            \shortstack{True \\ Positive} & 44 & 3 \\
            \shortstack{True \\ Negative} & 1 &  NA \\
        \end{tabular}
    \end{subtable}
    \begin{subtable}[t]{0.45\textwidth}
        \centering
        \caption{Nonmetastatic Lung Cancer}
        \begin{tabular}{c|c c}
            & \shortstack{Predicted \\ Positive} & \shortstack{Predicted \\ Negative} \\
            \hline
            \shortstack{True \\ Positive} & 57 & 1 \\
            \shortstack{True \\ Negative} & 2 &  NA \\
        \end{tabular}
    \end{subtable}

    \vspace{0.6cm}

    \begin{subtable}[t]{0.3\textwidth}
        \centering
        \caption{Oligometastatic Lung Cancer}
        \begin{tabular}{c|c c}
            & \shortstack{Predicted \\ Positive} & \shortstack{Predicted \\ Negative} \\
            \hline
            \shortstack{True \\ Positive} & 95 & 15 \\
            \shortstack{True \\ Negative} & 19 &  NA \\
        \end{tabular}
        \label{tab:conf_omd_lung}
    \end{subtable}
    \hfill
    \begin{subtable}[t]{0.3\textwidth}
        \centering
        \caption{Head and Neck Cancer}
        \begin{tabular}{c|c c}
            & \shortstack{Predicted \\ Positive} & \shortstack{Predicted \\ Negative} \\
            \hline
            \shortstack{True \\ Positive} & 117 & 17 \\
            \shortstack{True \\ Negative} & 13 &  NA \\
        \end{tabular}
        \label{tab:conf_head_neck}
    \end{subtable}
    \hfill
    \begin{subtable}[t]{0.3\textwidth}
        \centering
        \caption{Metastatic Melanoma}
        \begin{tabular}{c|c c}
            & \shortstack{Predicted \\ Positive} & \shortstack{Predicted \\ Negative} \\
            \hline
            \shortstack{True \\ Positive} & 70 & 18 \\
            \shortstack{True \\ Negative} & 13 &  NA \\
        \end{tabular}
        \label{tab:conf_melanoma}
    \end{subtable}
    \label{tab:conf}
\end{table}

\clearpage
\bibliographystyle{unsrt}
\bibliography{refs}

\end{document}